\documentclass[10pt]{article}
\usepackage{amsfonts}
\usepackage{amssymb,amsmath,amsthm,latexsym}
\newtheorem{theorem}{Theorem}[section]
\newtheorem{corollary}[theorem]{Corollary}
\newtheorem{proposition}[theorem]{Proposition}
\newtheorem{lemma}[theorem]{Lemma}
\newtheorem{definition}[theorem]{Definition}
\newtheorem{example}[theorem]{Example}

\newtheorem{remark}[theorem]{Remark}
\numberwithin{equation}{section}

\begin{document}
\title{Matrix Product Codes over Finite Commutative Frobenius Rings}

\author{Yun Fan${}^1$, San Ling${}^2$, Hongwei Liu${}^1$}
\date{${}^1$School of Mathematics and Statistics,
 Central China Normal University, Wuhan 430079, China \\
${}^2$Division of Mathematical Sciences, School of Physical \& Mathematical Sciences,
         Nanyang Technological University, Singapore 637616, Singapore}

\maketitle

\insert\footins{\small{\it Email addresses}: ~
 yunfan02@yahoo.com.cn (Yun Fan); ~
 lingsan@ntu.edu.sg (San Ling); ~
 h\_w\_liu@yahoo.com.cn (Hongwei Liu).
}

\begin{abstract}
Properties of matrix product codes over finite commutative Frobenius rings are investigated.
The minimum distance of matrix product codes constructed with several types of matrices
is bounded in different ways. The duals of matrix product codes
are also explicitly described in terms of matrix product codes.



\end{abstract}

\section{Introduction}

In coding theory, an interesting and important question is to
construct codes from smaller ones and to explore their properties via
those of the smaller ones. There have been many such
constructions, for example, the $(u|u + v)$-construction and the
$(a+x|b+x|a+b+x)$-construction. It was shown in \cite{LS-I} that 
quasi-cyclic codes over finite fields with co-index coprime
to the characteristic of the finite fields
can be constructed from linear codes of lower dimension
in a similar way, and the
$(a+x|b+x|a+b+x)$-construction is one such special case.
A more general construction, called the {\em matrix
product code}, which is formed by $m$ codes of length $n$ over a finite field
and an $m\times l$ matrix over the finite field,
was proposed and studied in \cite{Blackmore-Norton}.
Many, though not all, quasi-cyclic codes can be rewritten as matrix product codes,
for suitably chosen matrices.
It was further shown in \cite{Ozbudak-Stichtenoth} that the codes constructed
by algebraic geometry in \cite{Niederreiter-Xing} are in fact matrix product codes.
In~\cite{Blackmore-Norton}, a class of matrices, called {\em
non-singular by columns} matrices, was introduced, and some lower
bounds were obtained for the minimum distance of the matrix product codes
constructed with such matrices. However, most matrices
for quasi-cyclic codes, including the matrix for the
$(a+x|b+x|a+b+x)$-construction, are not non-singular by columns.
For general matrix product codes over finite fields,
a lower bound for the minimum distance was obtained in~\cite{Ozbudak-Stichtenoth}.
Decoding methods for some matrix product codes
were also discussed in \cite{H-H-R}, \cite{H-L-R} and \cite{H-R-11}. Other related work may 
be found in \cite{H-R-10}, \cite{M-M} and \cite{Ould-Mamoun}. 

On the other hand, coding over finite rings has attracted
much attention since the seminal work in \cite{Hammons}.
It was pointed out in the important works \cite{W-99} and \cite{W-08} that
only finite Frobenius rings are suitable for coding alphabets,
in the sense that several fundamental properties of codes over finite fields
still hold for codes over such rings. For example,
the double dual property, which says that the double dual coincides with
the original linear code, holds for linear codes over finite Frobenius rings.
A special class of finite Frobenius rings consists of the finite chain rings,
and codes over finite chain rings
have been investigated from many perspectives.
Recently, in \cite{Van-Asch}, matrix product codes over finite chain rings
were studied and the lower bound on the minimum distance of
matrix product codes by non-singular by columns matrices in \cite{Blackmore-Norton}
was extended to the minimum homogeneous distance.
Some quasi-cyclic codes over finite chain rings
have also been decomposed into matrix product codes
in \cite{LS-II}, though the terminology ``matrix product code''
was not used.

In this paper, we extend previous works on matrix
product codes in two directions. First, we formulate
matrix product codes over finite commutative Frobenius rings,
and explore their general properties, mainly,
the minimum distance and the structure of the duals.
Second, we consider new classes of matrices,
which contain the class of non-singular by columns
matrices as a special case, for which
we can bound the minimum distance of matrix product codes
thus constructed more precisely and more tightly,
and for which self-dual matrix product codes can be constructed efficiently.
The understanding of dual codes, as well as self-orthogonality and 
self-duality of codes, is a natural and important question in coding theory.

The organization of the paper is as follows.

Section 2 contains facts on matrices over finite commutative rings
which are needed for later sections, but which may not be readily available
in the literature.

In Section 3, we formulate matrix product codes
over finite commutative Frobenius rings,
and give two lower bounds for the minimum distance of such codes.
We also prove that the dual code of a matrix product code
is also a matrix product code whose structure is described precisely. Not
only does this extend earlier results in
\cite{Blackmore-Norton} and \cite{Van-Asch},
it also does not require the matrix to be a square matrix.

In Section 4, we introduce a class of matrices,
called {\em strongly full-row-rank (SFRR) matrices}
(see Definition \ref{Def4.1}),
which is bigger than the class of non-singular by columns matrices
and also contains certain matrices associated to quasi-cyclic codes.
We exhibit more precise lower bounds for the minimum distance of
matrix product codes constructed with these matrices,
as well as for their dual codes.
Besides extending corresponding results in~\cite{Blackmore-Norton},
conditions for which these lower bounds are attained are also given.

Inspired by the matrix for the $(a+x|b+x|a+b+x)$-construction, in Section~5
we consider special matrices, named
{\em two-way $(m')$-SFRR matrices} (see Definition \ref{Def5.1}),
and obtain lower and upper bounds
for the minimum distance of matrix product codes constructed with these matrices.
These bounds cover some known bounds for the minimum distance
of codes obtained from the $(a+x|b+x|a+b+x)$-construction as special cases.
For such matrices, we also show a condition (see Definition \ref{Def5.2})
which is useful for the construction of self-orthogonal matrix product codes.

\section{Matrices over Finite Commutative Rings}

In this paper, $R$ is always a finite commutative ring.
Writing the identity element~$1$ of the ring $R$ as the sum of
the primitive idempotents of $R$, we obtain an isomorphism
\begin{equation}\label{eq2.1}
 R\mathop{\longrightarrow}^{\cong}_\varphi R_1\oplus\cdots\oplus R_s,\quad
 r\longmapsto(r^{(1)},\cdots,r^{(s)}),
\end{equation}
where $R_1$, $\cdots$, $R_s$ are local commutative rings.
With the isomorphism (\ref{eq2.1}), in the following we usually identify
$R$ with $R_1\oplus\cdots\oplus R_s$ and just write
$r=(r^{(1)},\cdots,r^{(s)})$.

The finite commutative ring $R$ is called a {\em Frobenius ring}
if $R$ is self-injective (i.e., the regular module is injective),
or equivalently, $(C^{\bot})^\bot=C$ for any submodule $C$
of any free $R$-module $R^n$,
where $C^\bot$ denotes the orthogonal submodule of $C$
with respect to the usual Euclidean inner product on $R^n$.
Moreover, in this case, $|C^\bot||C|=|R|^n$ for any submodule $C$ of $R^n$,
where $|C|$ denotes the cardinality of $C$.
This is one of the reasons why only finite Frobenius rings
are suitable for coding alphabets.
With the isomorphism \eqref{eq2.1}, $R$ is Frobenius if and only if
every local component $R_i$ is Frobenius,
and the finite local commutative ring $R_i$ is Frobenius if and only if
$R_i$ has a unique minimal ideal.
Note that, in the non-commutative case, a self-injective ring
is called a {\em quasi-Frobenius ring}, while one more condition is required
for it to become a Frobenius ring. However, in the commutative case,
a finite quasi-Frobenius ring is exactly a finite Frobenius ring.
The reader may refer to \cite{W-99} for more details on Frobenius rings.

By ${\rm M}_{m\times l}(R)$, we mean the set of all $m\times l$ matrices over $R$.
For $A\in{\rm M}_{m\times l}(R)$, we denote the transpose of the matrix $A$ by $A^T$.
Given matrices $A$ of size $m \times l$ and $B$ of size $m \times l'$, we use
$(A|B)$ to denote the matrix of size $m \times (l + l')$ formed by concatenating $A$
and $B$. If $C$ is another matrix of size $m' \times l$,
the $(m + m') \times l$  matrix $\left( \begin{array}{c}
A \\ \hline C \end{array} \right)$
is similarly defined (by concatenating vertically).
We also let $0$ denote the zero matrix,
where the size will either be obvious from the context
or specified whenever necessary. Similarly,
we denote the $m \times m$ identity matrix by $I_m$,
or simply $I$ if the size is clear from the context.

Any matrix $A=(a_{ij})_{m\times l}\in{\rm M}_{m\times l}(R)$
can be written as
\begin{equation}\label{eq2.2}
A=\left(A^{(1)},\cdots,A^{(s)}\right)\,,\qquad
  A^{(k)}=\left(a_{ij}^{(k)}\right)_{m\times l}\in{\rm M}_{m\times l}(R_k),
  \quad 1\le k\le s,
\end{equation}
where the matrix addition and product are the coordinate-wise addition and product,
respectively.

Consider the free $R$-module $R^n$ of rank $n$.
Any element ${\bf a}=(a_1,\cdots,a_n)^T$
(written as a column vector) of $R^n$ is also called a vector, and we let ${\bf 0}$
denote the zero vector.
With the identification in (\ref{eq2.1}), we can write
$$
 R^n=R_1^n\oplus\cdots\oplus R_s^n,\quad
  {\bf a} = \big({\bf a}^{(1)},\cdots,{\bf a}^{(s)}\big),
$$
where ${\bf a}^{(k)}=(a_1^{(k)},\cdots,a_n^{(k)})^T$, for $1\le k\le s$,
is a column vector in $R_k^n$.

\begin{definition}\label{Def2.1}
For any integer $t \ge 1$, let ${\bf a}_i=(a_{i1},\cdots,a_{in})\in R^n$,
where $i=1,\cdots,t$.
The vectors ${\bf a}_1$,~$\cdots$,~${\bf a}_t$ are said to
be {\em linearly dependent} if there exists $(b_1,\cdots,b_t)$ in
the set difference $R^t \setminus \{ {\bf 0} \}$
such that $b_1{\bf a}_1+\cdots+b_t{\bf a}_t={\bf 0}$;
otherwise, ${\bf a}_1$, $\cdots$, ${\bf a}_t$ are said to be
{\em linearly independent}.
\end{definition}

If an $R$-submodule of $R^n$ is generated by vectors ${\bf a}_1$, $\cdots$, ${\bf a}_t$
which are linearly independent, then it is a free $R$-module of rank $t$
and we say that ${\bf a}_1$, $\cdots$, ${\bf a}_t$ form a {\em basis} of the free submodule.

The proof of the following result is straight-forward, so we omit it here. 

\begin{lemma}\label{Lem2.1}
The vectors ${\bf a}_1,\cdots,{\bf a}_t\in R^n$ are linearly
dependent if and only if there is an index $k$, with $1\le k\le s$, such
that ${\bf a}_1^{(k)},\cdots,{\bf a}_t^{(k)}\in R_k^n$ are
linearly dependent.
\end{lemma}

%

\begin{remark}\label{Rem-p4} {\rm The following is an equivalent formulation of
Lemma \ref{Lem2.1}:}

\noindent ``The vectors ${\bf a}_1,\cdots,{\bf a}_t\in R^n$ are linearly
independent if and only if, for all $k$ with $1\le k\le s$, the vectors
${\bf a}_1^{(k)},\cdots,{\bf a}_t^{(k)}\in R_k^n$ are linearly independent.''
\end{remark}

\begin{definition}\label{Def2.2}
Let $A=(a_{ij})_{m\times l}$ be a matrix over $R$.

\begin{itemize}
\item[(i)] If the rows of $A$ are linearly independent,
then we say that $A$ is a {\em full-row-rank (FRR)} matrix.

\item[(ii)] If there is an $l\times m$ matrix $B$ over $R$ such that
$AB=I$, then we say that
$A$ is {\em right-invertible} and $B$ is a {\em right inverse} of $A$.

\item[(iii)] If $m=l$ and the determinant $\det A$ is a unit of $R$,
      then we say that $A$ is {\em non-singular}.

\item[(iv)] If, for every $t$ with $1\le t\le m$, any $t\times t$ submatrix
of the first (resp., last) $t$ rows of $A$ is non-singular,
then we say that $A$ is {\em non-singular by columns} (resp., {\em reversely
non-singular by columns}).
\end{itemize}
\end{definition}

\begin{remark}\label{rem-p4-new}
{\rm
\begin{itemize}
\item[(i)] It is obvious that, if $A$ is a matrix over $R$ of size $m\times l$,
and $P$, $Q$ are invertible matrices over $R$ of size $l\times l$ and $m\times m$, respectively,
then $A$, $AP$ and $QA$ are all FRR provided one of them is FRR.

\item[(ii)] By Remark \ref{Rem-p4}, a matrix $A$ over $R$ is FRR if and only if
the matrices $A^{(k)}$ over $R_k$ in (\ref{eq2.2}), for $k=1,\cdots,s$, are all FRR.
\end{itemize} }
\end{remark}

As in usual linear algebra, the following two types of operations
are called elementary row (or column) operations on matrices over $R$:

\begin{itemize}
\item adding a multiple of a row (column) to another row (column),

\item multiplying a row (column) by a unit of $R$.
\end{itemize}

\begin{lemma}\label{Lem2.2}
Assume that $R$ is a finite local ring
and~$A=\left(a_{ij}\right)_{m\times l}$ is a matrix over $R$. Then
$A$ is FRR if and only if $m\le l$ and
there is an invertible $l\times l$ matrix $P$ over $R$ such that
$AP=\left(\,I\mid 0\,\right)_{m\times l}$. In particular,
$A$ is FRR if and only if $A$ is right invertible.
\end{lemma}

\noindent{\bf Proof.}~ Note that $R$ has a unique maximal ideal $J$
such that the set difference $R \setminus J$ is just the set of all units of $R$.
Since $R$ is finite, there is an integer $e>0$ such that $J^e=0$ but
$J^{e-1}\ne 0$ ($e$ is called the {\em nilpotency index} of $J$,
and we adopt the convention that $e=1$ if $R$ is a field).
Thus we can pick a $\delta\in J^{e-1}$ with $\delta\ne 0$.
For any row $(a_{i1},\cdots,a_{il})$ of $A$, we claim that

$\bullet$~ {\it There is an entry $a_{ij}$ which is a unit of $R$.}

\noindent For, otherwise, all $a_{i1},\cdots,a_{il}$ belong to $J$ and
hence all $\delta a_{i1},\cdots,\delta a_{il}$ belong to $J^e=\{0\}$,
that is, $\delta\cdot\left(a_{i1},\cdots,a_{il}\right)= {\bf 0}$, and
the row $(a_{i1},\cdots,a_{il})$ of $A$ is linearly dependent,
which contradicts the assumption that $A$ is FRR.

Therefore, in the first row of $A$, we can find a unit.
After some suitable permutation of the columns, we can assume that $a_{11}$ is a unit.
With appropriate elementary operations on the columns,
we can transform $A$ into an FRR matrix as follows:
$$ \begin{pmatrix} 1 & 0 & \cdots & 0 \\
        a'_{21} & a'_{22} & \cdots & a'_{2l} \\
        \cdots & \cdots & \cdots & \cdots  \\
        a'_{m1}& a'_{m2} & \cdots & a'_{ml}\end{pmatrix}.$$
Next we assert that

$\bullet$~ {\it Some $a'_{2j}$, for $2\le j\le l$, is a unit of $R$.}

\noindent Assuming the contrary, then
$\delta a'_{21}\cdot(1,0,\cdots,0)-\delta(a'_{21},a'_{22},\cdots,a'_{2l})={\bf 0}$,
which contradicts the assumption that the above matrix is FRR.

One can continue with elementary operations on the columns in the same manner,
until the desired form $\left(\,I\mid 0\,\right)$ is obtained. \qed

\medskip Now we return to the general case where $R$ may be not local,
and we identify $R$ with the direct sum
$R_1\oplus\cdots\oplus R_s$ of local Frobenius rings $R_k$, $k=1,\cdots,s$,
by the isomorphism (\ref{eq2.1}). Then we obtain the following:

\begin{corollary}\label{Cor2.3}
$A\in{\rm M}_{m\times l}(R)$ is FRR if and only if $A$ is
right-invertible.
\end{corollary}

\noindent{\bf Proof.}~ By Lemma \ref{Lem2.1}, the matrix $A$ over $R$
is FRR if and only if every $A^{(k)}$ over $R_k$,
for $k=1,\cdots,s$, is FRR (see (\ref{eq2.2})).
Further, by Lemma \ref{Lem2.2}, for $k=1,\cdots,s$, every $A^{(k)}\in{\rm M}_{m\times l}(R_k)$,
is FRR if and only if there is
$B^{(k)}\in{\rm M}_{l\times m}(R_k)$ such that $A^{(k)}B^{(k)}=I$.
Setting $B=\left(B^{(1)},\cdots,B^{(s)}\right)\in{\rm M}_{l\times m}(R)$,
we obtain $AB=I$. \qed

The following corollary follows from a typical linear algebra argument.

\begin{corollary}\label{Cor2.4}
Let $A$ be in ${\rm M}_{m\times m}(R)$. The following statements are equivalent:
\begin{itemize}
\item[(i)] $A$ is invertible.

\item[(ii)] $A$ is non-singular.

\item[(iii)] $A$ is FRR. 
\end{itemize}
\end{corollary}

\begin{proposition}\label{Prop2.5}
Let $A\in{\rm M}_{m\times l}(R)$
be FRR and let $X=(x_1,\cdots,x_l)^T$, where $x_i$'s are variables.
Then the set of solutions of the linear equation system $AX={\bf 0}$
is a free submodule in $R^l$ of rank $l-m$ and we have an
FRR $(l-m)\times l$ matrix $G$ over $R$
whose rows form a basis of this free submodule.
\end{proposition}

\noindent{\bf Proof.}~ First, assume that $R$ is local.
By Lemma \ref{Lem2.2}, we have an invertible matrix $P$ of size $l\times l$ such that
$AP=\left(\,I\mid 0\,\right)_{m\times l}$.
The set of solutions of the linear equation system
$(AP)Y={\bf 0}$ in variables $Y=(y_1,\cdots,y_l)^T$
is clearly a free submodule of $R^l$ of rank $l-m$ with the rows of
the matrix $\left(\,0\mid I\,\right)_{(l-m)\times l}$ as a basis.
Rewriting $AX={\bf 0}$ as $(AP)(P^{-1}X)={\bf 0}$, we see that
the set of solutions of $AX={\bf 0}$ is a free submodule of $R^l$ of rank $l-m$
with the rows of the matrix $G=\left(\,0\mid I\,\right)_{(l-m)\times l}P^T$
as a basis.

Returning to the general case where $R$ is a commutative Frobenius ring,
we have the identification in (\ref{eq2.1}).
For each index $1\le k\le s$, we have
a linear equation system $A^{(k)}X^{(k)}={\bf 0}$ with the matrix $A^{(k)}$
over the local ring $R_k$ being FRR (see Lemma \ref{Lem2.1}),
so we have an FRR matrix $G^{(k)}$ over $R_k$ of size $(l-m)\times l$
such that the rows of $G^{(k)}$ form a basis of the free submodule of $R_k^l$
of the solutions of the system $A^{(k)}X^{(k)}={\bf 0}$.
With the identification (\ref{eq2.2}), we can construct a matrix
$G=\left(G^{(1)},\cdots,G^{(s)}\right)$ over $R$ of size $(l-m)\times l$
which is FRR too, and any vector ${\bf a}\in R^l$ is a solution
of the system $AX={\bf 0}$ if and only if
${\bf a}$ is a combination of the rows of $G$.
In other words, the set of solutions of the system $AX={\bf 0}$ is a free submodule
of $R^l$ of rank $l-m$ with the rows of $G$ as a basis. \qed

\begin{remark}\label{Rem-p6}
{\rm With $A,G$ as in Proposition \ref{Prop2.5}, denote by $L_G$ and $L_A$ the free submodules
of $R^l$ generated by the rows of $G$ and $A$, respectively.
With the usual Euclidean inner product $\langle -,-\rangle$ on~$R^l$,
Proposition \ref{Prop2.5} says that $(L_A)^\bot=L_G$. As a consequence,
we see that}
\begin{itemize}
\item If $R$ is a finite commutative Frobenius ring, then a submodule $V$ of $R^l$
is free if and only if its orthogonal submodule $V^\bot$ is free.
\end{itemize}
{\rm
The ``only if'' part is just Proposition \ref{Prop2.5}. For the ``if'' part,
taking a generator matrix~$A$ of~$V^\bot$
(i.e., $A$ is FRR and $V^\bot=L_A$), since $R$ is a Frobenius ring,
we have that $V=(V^\bot)^\bot=(L_A)^\bot=L_G$ is free.
}
\end{remark}

\begin{proposition}\label{Prop2.6}
Any FRR $m\times l$ matrix $A$ over $R$ can be, by appending rows,
extended to an invertible $l\times l$ matrix
$\tilde A=\left(\begin{array}{c}A\\ \hline A'\end{array}\right)$
(equivalently, any set of linearly independent vectors of $R^l$
can be extended to a basis of $R^l$).
Furthermore, for any such extension
$\tilde A=\left(\begin{array}{c}A\\ \hline A'\end{array}\right)$,
partitioning $\tilde A^{-1}=(B\,|\,B')$ into
an $l\times m$ submatrix $B$ and an $l\times(l-m)$ submatrix $B'$,
we have that $B$ is a right inverse of $A$
and $B'^T$ is a generator matrix of the submodule
of solutions of the linear equation system $AX={\bf 0}$.
\end{proposition}

\noindent{\bf Proof.}~
By Lemma \ref{Lem2.2}, we have a right inverse $B$ of $A$,
and we denote by $B_1,\cdots,B_m$ the columns of $B$.
By Proposition \ref{Prop2.5},
we have an $(l-m)\times l$ matrix $G$ whose rows form a basis
of the free submodule of solutions of the linear equation system $AX={\bf 0}$,
and we denote by $G_1^T,\cdots,G_{l-m}^T$ the columns of $G^T$.
Then we form an $l\times l$ matrix $\tilde B=(B\,|\,G^T)$.
Suppose $d_1,\cdots,d_m,e_1,\cdots,e_{l-m}\in R$ such that
\begin{equation}\label{eq-p6}
 d_1B_1+\cdots+d_m B_m +e_1G_1^T+\cdots+e_{l-m}G_{l-m}^T={\bf 0}.
\end{equation}
Then, since $G_i^T$'s are solutions of $AX={\bf 0}$, we have
$$
{\bf 0}=d_1AB_1+\cdots+d_m AB_m +e_1AG_1^T+\cdots+e_{l-m}AG_{l-m}^T
   =d_1AB_1+\cdots+d_m AB_m.
$$
However, since $AB=I$, we get that $d_1=\cdots=d_m=0$.
Returning to (\ref{eq-p6}),
we have that $e_1G_1^T+\cdots+e_{l-m}G_{l-m}^T={\bf 0}$,
hence $e_1=\cdots=e_{l-m}=0$ since $G$ is FRR.
Thus, $\tilde B$ is a square matrix with linearly independent columns and it
is hence invertible. Expressing $\tilde B^{-1}$ as
$\tilde B^{-1}=\left(\begin{array}{c}A''\\\hline A'\end{array}\right)$,
where $A''$ and $A'$ are formed by the first $m$ and the last
$l-m$ rows, respectively, of $\tilde B^{-1}$, we can rewrite
$\tilde B^{-1}\tilde B=I$ as
$$\left(\begin{array}{c}A''\\\hline A'\end{array}\right)\cdot
 (B\,|\,G^T)=\begin{pmatrix}A''B& A''G^T\\ A'B&A'G^T \end{pmatrix}
 =\begin{pmatrix}I&0\\0&I\end{pmatrix}.$$
In particular, $A'\cdot(B\,|\,G^T)=(0\,|\,I)$.
On the other hand, it follows from our choices of $B$ and $G$
that $A\cdot(B\,|\,G^T)=(I\,|\,0)$. Therefore,
\begin{equation}\label{eq-p7}
\left(\begin{array}{c}A\\\hline A'\end{array}\right)\cdot
 (B\,|\,G^T)=\begin{pmatrix}AB& AG^T\\ A'B&A'G^T \end{pmatrix}
 =\begin{pmatrix}I&0\\0&I\end{pmatrix}.\end{equation}
Thus $\left(\begin{array}{c}A\\\hline A'\end{array}\right)$
is right invertible, which, by Corollary \ref{Cor2.4}, means that it is invertible
and $(B\,|\,G^T)$ is an inverse of it.

Similar to the equality (\ref{eq-p7}), for any $(l-m)\times l$ matrix $A'$,
$l\times m$ matrix $B$ and $l\times(l-m)$ matrix~$B'$, the equality
$\left(\begin{array}{c}A\\\hline A'\end{array}\right)\cdot
 (B\,|\,B')=\begin{pmatrix}I&0\\0&I\end{pmatrix}$
implies that $AB=I$ and $AB'=0$.
\qed

\section{Matrix Product Codes over Frobenius Rings}

Starting from this section till the end of this paper, we assume that $R$ 
is always a finite commutative Frobenius ring as in (\ref{eq2.1}).

Any non-empty subset $C$ of $R^n$ is called a code over $R$ of length $n$ and
any vector in $C$ is called a codeword. Let $M$ denote the cardinality of $C$,
i.e., $M=|C|$. Then $C$ is said to be an $(n,M)$ code over $R$.
If $C$ is an $R$-submodule of $R^n$, then $C$ is called a linear code.
With respect to the usual Euclidean inner product,
we have the dual code $C^\bot$ which is always linear. When 
$C \subseteq C^\bot$ (resp., $C = C^\bot$), we say that $C$ is self-orthogonal 
(resp., self-dual).
If $C$ is linear, then $(C^{\bot})^{\bot}=C$ and $|C|\cdot|C^\bot|=|R|^n$,
as we have noted in Section 2.

Let $A=(a_{ij})_{m\times l}\in{\rm M}_{m\times l}(R)$.
For any index $1\le k\le m$, we denote by $U_A(k)$ the linear
code over $R$ of length $l$ generated by the $i$th rows of $A$,
for $i=1,2,\cdots,k$,
and denote by $L_A(k)$ the linear code over $R$ of length $l$ generated by
the $i$th rows of $A$, for $i=k,k+1,\cdots, m$.
In particular, $U_A(m)=L_A(1)$ is the linear code over $R$ of length $l$
generated by all the rows of $A$.
Thus, the set of solutions of the linear equation system $AX={\bf 0}$
is just the dual code $L_A(1)^\bot$ of the code $L_A(1)$.
If $A$ is FRR, then $L_A(1)$ is a
free submodule of $R^l$ of rank $m$, while its dual $L_A(1)^\bot$
is a free submodule of $R^l$ of rank $l-m$, and the matrix $G$ in Proposition \ref{Prop2.5}
is a generator matrix of $L_A(1)^\bot$,
i.e., $L_A(1)^\bot=L_G(1)$. For convenience, we also define $U_A(0)$ and
$L_A(m+1)$ to be the zero code.

Any $n\times m$ matrix can be viewed as a word over $R$ of length $nm$,
so any non-empty subset $D$ of ${\rm M}_{n\times m}(R)$ can be viewed as a code
over $R$ of length $nm$. From this point of view, for any two words
${\bf w},{\bf v}\in{\rm M}_{n\times m}(R)$,
the Euclidean inner product can be computed as follows
\begin{equation}\label{eq3.1}
 \langle{\bf w},{\bf v}\rangle={\rm tr}({\bf w}{\bf v}^T) ,
\end{equation}
where ${\rm tr}({\bf w}{\bf v}^T)$ denotes the trace of the
$n\times n$ matrix ${\bf w}{\bf v}^T$. For: writing ${\bf
w}=(w_{ij})_{n\times m}$, ${\bf v}=(v_{ij})_{n\times m}$, then
${\rm tr}({\bf w}{\bf v}^T)=\sum_{i=1}^n\sum_{j=1}^m
w_{ij}v_{ij}$, which is just the Euclidean inner product of ${\bf
w}$ and ${\bf v}$. Note that \eqref{eq3.1} holds for any
matrix size, including the usual words written in the form of row or column vectors.

Let $A$ be an FRR $m\times l$ matrix over $R$, then the map
$$
 {\rm M}_{n\times m}(R) \longrightarrow {\rm M}_{n\times l}(R),\quad
 {\bf v}\longmapsto {\bf v}A
$$
is an injective linear map, for: $A$ has a right inverse $B$, so that,
if ${\bf d}A={\bf d'}A$, then ${\bf d}={\bf d}I={\bf d}AB={\bf d'}AB={\bf d'}$.
Therefore, if the subset $D$ of ${\rm M}_{n\times m}(R)$ is an $(nm,M)$ code over $R$,
then $DA=\left\{{\bf d}A\mid{\bf d}\in D\right\}$ is an $(nl, M)$ code over $R$,
and $DA$ is linear if and only if $D$ is linear.

Let $C_j$ be an $(n,M_j)$ code over $R$, for $j=1,\cdots,m$.
For ${\bf c}_1\in C_1,$ $\cdots$, ${\bf c}_m\in C_m$, we have an $n\times m$ matrix
$({\bf c}_1,\cdots,{\bf c}_m)$, where each ${\bf c}_j$ is written as a column vector.
Hence, we have a subset of ${\rm M}_{n\times m}(R)$ as follows:
$$ D=[C_1,\cdots,C_m]=\left\{({\bf c}_1,\cdots,{\bf c}_m)\mid
 {\bf c}_1\in C_1,\cdots,{\bf c}_m\in C_m \right\}. $$
Obviously, $[C_1,\cdots,C_m]$ is an $\left(nm,\prod_{j=1}^mM_j\right)$ code over $R$,
and the code $[C_1,\cdots,C_m]$ is linear
if and only if all $C_1,\cdots,C_m$ are linear.

Let $A$ be an FRR $m\times l$ matrix over $R$. We have
an $\left(n l,\prod_{j=1}^mM_j\right)$ code over $R$, called a
{\em matrix product code} over $R$ (see \cite{Blackmore-Norton}),
as follows:
\begin{equation}\label{eq3.2}
 [C_1,\cdots,C_m]A=\left\{({\bf c}_1,\cdots,{\bf c}_m)A\mid
 {\bf c}_1\in C_1,\cdots,{\bf c}_m\in C_m \right\},
\end{equation}
which is linear if all $C_1,\cdots,C_m$ are linear.

It is easy to check that $[C_1,\cdots,C_m]A=[C_1,\cdots,C_m]$
if $C_1,\cdots,C_m$ are all linear, $A$ is square, and one of the following holds:
\begin{itemize}
\item $A$ is a diagonal matrix,

\item  $C_1\supseteq C_2\supseteq\cdots\supseteq C_m$ and
$A$ is a lower triangular matrix,

\item $C_1=C_2=\cdots=C_m$.
\end{itemize}

Any weight $w$ on $R$ can be extended to
a weight on $R^n$ in the obvious way,
hence the distance $d_w$ on $R^n$ with respect to the weight $w$ is defined by
$d_w({\bf c},{\bf c'})=w({\bf c}-{\bf c'})$ for ${\bf c},{\bf c'}\in R^n$.
The minimum distance of any code $C$ with respect to the weight $w$,
denoted by $d_w(C)$, is defined to be the minimum distance
with respect to the weight $w$ between any two distinct codewords in $C$;
and we adopt the convention that $d_w(0)=n+1$
for the zero code $0=\{{\bf 0}\} \subseteq R^n$.
In particular, we denote the Hamming weight by $w_H$
and the Hamming distance by $d_H$,
hence $d_H(C)$ denotes the minimum Hamming distance of $C$.

The following is a generalization of the main result of \cite{Ozbudak-Stichtenoth}
to matrix product codes over finite Frobenius rings.

\begin{theorem}\label{Thm3.1}
Let $C_j$ be an $(n,M_j)$ code over $R$, for $j=1,\cdots,m$, and let
$A=(a_{ij})_{m\times l}$ be an FRR matrix over $R$. Let
$w$ be a weight on $R$. Then
$C=[C_1,\cdots,C_m]A$ is an $\left(nl, \prod_{j=1}^m M_j\right)$ code
over $R$ with minimum distance $d_w(C)$ satisfying
\stepcounter{equation}
\begin{equation}\tag{\theequation U}\label{eq3.3U}
d_w(C)\ge \min\left\{d_H(C_k)d_w\big(U_A(k)\big)\mid k=1,\cdots,m\right\},
\end{equation}
\begin{equation}\tag{\theequation L}\label{eq3.3L}
d_w(C)\ge \min\left\{d_H(C_k)d_w\big(L_A(k)\big)\mid k=1,\cdots,m\right\}.
\end{equation}
\end{theorem}

\noindent{\bf Proof.}~ Since $A$ is FRR, by \eqref{eq3.2} we have that
 $C$ is an $\big(nl, \prod_{j=1}^m M_j\big)$ code over $R$.

For any two distinct codewords ${\bf c}=({\bf c}_1,\cdots,{\bf c}_m)A$,
${\bf c}'=({\bf c}'_1,\cdots,{\bf c}'_m)A$ of $C$, let
${\bf c}_j-{\bf c}'_j={\bf b}_j$, for $j=1,\cdots,m$.
Then ${\bf c}-{\bf c}'=({\bf b}_1,\cdots,{\bf b}_m)A$ and
$d_w({\bf c},{\bf c}')=w({\bf c}-{\bf c}')=w\big(({\bf b}_1,\cdots,{\bf b}_m)A\big)$.
Note that there is an index $k$ such that ${\bf b}_j={\bf 0}$ for all $j<k$
but ${\bf b}_k\ne{\bf 0}$. Let $A_i$ denote the $i$th row of $A$.
Then the word ${\bf c}-{\bf c}'$ which is an $n\times l$ matrix over $R$ is as follows:
$$
{\bf c}-{\bf c}'=({\bf 0},\cdots,{\bf 0}, {\bf b}_k,\cdots,{\bf b}_m)A
 = ({\bf b}_k,\cdots,{\bf b}_m)\begin{pmatrix}A_{k}\\ \vdots\\
 A_m\end{pmatrix},
$$
where ${\bf b}_k=(b_{1k},\cdots,b_{ik},\cdots,b_{nk})^T$ with $b_{ik}\in R$.
For each non-zero $b_{ik}$, we get the $i$th row of the matrix ${\bf c}-{\bf c}'$ as follows:
$$
 b_{ik} A_k + b_{i,k+1}A_{k+1}+\cdots+b_{im}A_m,
$$
which is a non-zero codeword of the code $L_A(k)$.
Therefore, the contribution to $d_w({\bf c},{\bf c}')$ of the $i$th row of ${\bf c}-{\bf c}'$
is $w(b_{ik} A_k + b_{i,k+1}A_{k+1}+\cdots+b_{im}A_m)\ge d_w(L_A(k))$.
Since $w_H({\bf b}_k)=d_H({\bf c}_k,{\bf c}'_k)$,
the number of non-zero $b_{ik}$ is at least $d_H(C_k)$. In conclusion,
$d_w({\bf c},{\bf c}')\ge d_H(C_k)d_w(L_A(k))$.
Thus the inequality \eqref{eq3.3L} holds.

Similarly, for ${\bf c}, {\bf c}'$ above, there is an index $k'$ such that
${\bf b}_j={\bf 0}$ for all $j>k'$ but ${\bf b}_{k'}\ne{\bf 0}$, so we can write
${\bf c}-{\bf c}'$ as follows:
$$
{\bf c}-{\bf c}'=({\bf b}_1,\cdots,{\bf b}_{k'},{\bf 0},\cdots,{\bf 0})A
 = ({\bf b}_1,\cdots,{\bf b}_{k'})\begin{pmatrix}A_{1}\\ \vdots\\ A_{k'}\end{pmatrix},
$$
and obtain that $d_w({\bf c},{\bf c}')\ge d_H(C_{k'})d_w(U_A(k'))$.
We are done for the inequality \eqref{eq3.3U}. \qed

\begin{remark}\label{Rem-p9}
{\rm
\begin{itemize}
\item[(i)] Though, in the above proof, it is stated:
``there is an index $k$ such that ...'',
in fact any index $k$ can appear when ${\bf c},{\bf c}'$ run over the choices of
two distinct codewords of~$C$, since we can choose ${\bf c}_j={\bf c}'_j$, for $j\ne k$,
and ${\bf c}_k\ne{\bf c}'_k$.

\item[(ii)] In general,
the right hand sides of \eqref{eq3.3U} and \eqref{eq3.3L} are not strict lower bounds
of the minimum distance (see Section 5).

\item[(iii)] The two lower bounds in \eqref{eq3.3U} and \eqref{eq3.3L} cannot be 
directly compared
in general: sometimes \eqref{eq3.3U} is better than \eqref{eq3.3L},
while some other times the opposite is true.
\end{itemize} }
\end{remark}

The following result describes the dual of a matrix product code constructed with an FRR matrix.
It may be regarded as a generalization of \cite[Theorem 6.6]{Blackmore-Norton}
and \cite[Proposition 3]{Van-Asch}, but here we do not require the matrix to be square.

\begin{theorem}\label{Thm3.2}
Let $C_1, \cdots , C_m$ be codes over $R$ of length $n$, and
let $A\in{\rm M}_{m\times l}(R)$ be FRR.
Assume that $B\in{\rm M}_{l\times m}(R)$ is a right inverse of $A$
and $G\in{\rm M}_{(l-m)\times l}(R)$ is a generator matrix of the
dual code $L_A(1)^\bot$ of $L_A(1)$.
Set $\tilde B=\big(B\,|\,G^T\big)$.
Then the dual code of $C=[C_1,\cdots,C_m]A$ is
\begin{equation}\label{eq3.4}
C^\bot=[\,C_1^\bot,\cdots,C_m^\bot,\,\underbrace{R^n,\cdots,R^n}_{l-m}\,]\tilde B^T
   =[C_1^\bot,\cdots,C_m^\bot]B^T+{\rm M}_{n\times(l-m)}(R)G.
\end{equation}
\end{theorem}

\noindent{\bf Proof.}~
We denote by $\hat C_j$ the linear code generated by the vectors in $C_j$,
and by $\hat C$ the linear code generated by the vectors in $C$.
It is then easy to check that $C_j^\bot=\hat C_j^\bot$,
$\hat C=[\hat C_1,\cdots,\hat C_m]A$, and $C^\bot=\hat C^\bot$.
Thus, without loss of generality, in the following we assume
that $C_1,\cdots,C_m$ are all linear codes.

In the equality (\ref{eq-p7}) within the proof of Proposition \ref{Prop2.6},
we have seen that $\tilde B=\big(B\,|\,G^T\big)$ is an
invertible $l\times l$ matrix such that~$A$ is the $m\times l$ submatrix
of $\tilde A=\tilde B^{-1}$ formed by the first $m$ rows of $\tilde A$,
i.e., $\tilde A=\tilde B^{-1}$ is partitioned as
$\tilde A=\left(\begin{array}{c} A\\\hline A'\end{array}\right)$.
It is obvious that
\begin{equation}\label{eq3.5}
 C=[C_1,\cdots,C_m]A=[C_1,\cdots,C_m,\,\underbrace{0,\cdots,0}_{l-m}\,]\tilde A .
\end{equation}
Now we show that
\begin{equation}\label{eq3.6}
[C_1^\bot,\cdots,C_m^\bot,\,\underbrace{R^n,\cdots,R^n}_{l-m}\,]\tilde B^T
  ~\subseteq~ C^\bot.
\end{equation}
Let ${\bf c}=({\bf c}_1,\cdots,{\bf c}_m,{\bf 0},\cdots,{\bf 0})\tilde A\in C$
with ${\bf c}_j\in C_j$, and let
${\bf d}=({\bf d}_1,\cdots,{\bf d}_m,{\bf w}_{m+1},\cdots,{\bf w}_l)\tilde B^T$
with ${\bf d}_j\in C_j^\bot$ ($1 \le j \le m$) and ${\bf w}_j\in R^n$ ($m+1 \le j \le l$).
By \eqref{eq3.1}, we have
\begin{eqnarray*}
\langle{\bf c},{\bf d}\rangle&=&{\rm tr}
 \left(({\bf c}_1,\cdots,{\bf c}_m,{\bf 0},\cdots,{\bf 0})\tilde A\cdot
  \big(({\bf d}_1,\cdots,{\bf d}_m,{\bf w}_{m+1},\cdots,{\bf w}_l)\tilde B^T\big)^T\right)\\
 &=&{\rm tr}\left(({\bf c}_1,\cdots,{\bf c}_m,{\bf 0},\cdots,{\bf 0})\tilde A
  \tilde B\begin{pmatrix}{\bf d}_1^T\\ \vdots\\{\bf d}_m^T\\
    {\bf w}_{m+1}^T\\\vdots\\{\bf w}_{l}^T \end{pmatrix}
  \right).
\end{eqnarray*}
Since $\tilde A\tilde B=I$ is the identity matrix, we obtain
$$\langle{\bf c},{\bf d}\rangle
 ~=~{\rm tr}\left(({\bf c}_1,\cdots,{\bf c}_m)
  \begin{pmatrix}{\bf d}_1^T\\ \vdots\\{\bf
  d}_m^T\end{pmatrix}\right).
$$
By the linearity of trace, we have
$$\langle{\bf c},{\bf d}\rangle
  ={\rm tr}\left({\bf c}_1{\bf d}_1^T+\cdots+{\bf c}_m{\bf d}_m^T\right)
  ={\rm tr}\left({\bf c}_1{\bf d}_1^T\right)+\cdots
   +{\rm tr}\left({\bf c}_m{\bf d}_m^T\right).
$$
By \eqref{eq3.1} again, we obtain
$$\langle{\bf c},{\bf d}\rangle=
\langle{\bf c}_1,{\bf d}_1\rangle+\cdots+\langle{\bf c}_m,{\bf d}_m\rangle=0.
$$
Thus \eqref{eq3.6} is proved.

Since $R$ is a Frobenius ring, $|C_j^\bot|=\frac{|R|^n}{|C_j|}$,
for $j=1,\cdots,m$, and $|C^\bot|=\frac{|R|^{nl}}{|C|}$.
It follows from \eqref{eq3.2} that
\begin{eqnarray*}
 \big|[C_1^\bot,\cdots,C_m^\bot,\,\underbrace{R^n,\cdots,R^n}_{l-m}\,]\tilde B^T\big|
 &=&|C_1^\bot|\cdots|C_m^\bot|\cdot\,\underbrace{|R^n|\cdots|R^n|}_{l-m}\\
 &=&\frac{|R|^n}{|C_1|}\cdots\frac{|R|^n}{|C_m|}\cdot|R|^{n(l-m)}=\frac{|R|^{nl}}{|C|}
 =|C^\bot|\,.
\end{eqnarray*}
Therefore, the equality in \eqref{eq3.6} must hold. In other words, we obtain
$$ C^\bot=[\,C_1^\bot,\cdots,C_m^\bot,\,\underbrace{R^n,\cdots,R^n}_{l-m}\,]\tilde B^T, $$
which is the first equality in \eqref{eq3.4}.

Further, since $\tilde B^T$ has the partitioned form
$\tilde B^T=\left(\begin{array}{c} B^T\\\hline G\end{array}\right)$,
\begin{eqnarray*}
[C_1^\bot,\cdots,C_m^\bot,\,\underbrace{R^n,\cdots,R^n}_{l-m}\,]\tilde B^T
 &=&[C_1^\bot,\cdots,C_m^\bot]B^T+[\,\underbrace{R^n,\cdots,R^n}_{l-m}\,]G\\
 &=&[C_1^\bot,\cdots,C_m^\bot]B^T+{\rm M}_{n\times(l-m)}(R)G ,
\end{eqnarray*}
i.e., the second equality in \eqref{eq3.4} holds.
\qed

\begin{remark}\label{Rem-p10}
{\rm By Proposition \ref{Prop2.6}, the conclusion of Theorem \ref{Thm3.2}
can be rewritten as follows: for any $l\times l$ matrix
$\tilde A=\left(\begin{array}{c} A\\\hline A'\end{array}\right)$
and $\tilde A^{-1}=\big(B\,|\,B'\big)$, we have that
$$
C^\bot=[\,C_1^\bot,\cdots,C_m^\bot,\,\underbrace{R^n,\cdots,R^n}_{l-m}\,]
 (\tilde{A}^{-1})^T
   =[C_1^\bot,\cdots,C_m^\bot]B^T+{\rm M}_{n\times(l-m)}(R)B'^T.
$$ }
\end{remark}

\medskip
An $m\times l$ matrix $A$ over $R$, where $m\le l$,
is said to be {\em quasi-orthogonal} if $AA^T$ is a diagonal square matrix
where all the diagonal entries are units of $R$.
For example, the matrix $\begin{pmatrix}1&1&1&0\\0&1&1&1\end{pmatrix}$
is quasi-orthogonal if the characteristic of $R$ is $2$, while
the matrix $\begin{pmatrix}1&1&0\\0&0&1\end{pmatrix}$
is quasi-orthogonal if the characteristic of $R$ is $3$.

\begin{theorem}\label{Thm3.3}
Let $C_1,\cdots,C_m$ be self-orthogonal linear codes over $R$ of length $n$,
let $A$ be a quasi-orthogonal $m\times l$ matrix over $R$
and let $G$ be a generator matrix of the dual code $L_A(1)^\bot$ of $L_A(1)$.
Then the dual code of $C=[C_1,\cdots,C_m]A$ is
$C^\bot=[C_1^\bot,\cdots,C_m^\bot]A+{\rm M}_{n\times(l-m)}(R)G$.
In particular, $C$ is a self-orthogonal code.
\end{theorem}

\noindent{\bf Proof.}~
Assume that $AA^T=D=\begin{pmatrix}u_1\\ &\ddots\\&& u_m\end{pmatrix}$,
with $u_i$ being units of $R$.
Then $A^TD^{-1}$ is a right inverse of $A$.
By the equality (\ref{eq-p7}) in the proof of Proposition \ref{Prop2.6},
the matrix $\left( A^TD^{-1} \vert G^T\right)$ is invertible,
hence $\left( A^T \vert G^T\right)=\left( A^TD^{-1} \vert G^T\right)
\begin{pmatrix}D \\ & I\end{pmatrix}$ is invertible.
Thus, $\left(\begin{array}{c} A\\ \hline G \end{array}\right)
=\left( A^T \vert G^T\right)^T$ and the product
\begin{equation}\label{eq3.7}
 \left(\begin{array}{c} A\\ \hline G \end{array}\right)
  \left( A^T \vert G^T\right)=\begin{pmatrix} D\\ & GG^T \end{pmatrix}
\end{equation}
are invertible; hence $GG^T$ is an invertible
$(l-m)\times(l-m)$ matrix, and
$\left( A^T \vert G^T\right)\begin{pmatrix} D^{-1}\\ & (GG^T)^{-1}\end{pmatrix}$
is the inverse of $\left(\begin{array}{c} A\\ \hline G \end{array}\right)$.
Note that
$$
\left(\left( A^T \vert G^T\right)\begin{pmatrix} D^{-1}\\ & (GG^T)^{-1}\end{pmatrix}\right)^T
=\begin{pmatrix} D^{-1}\\ & (GG^T)^{-1}\end{pmatrix}
 \left(\begin{array}{c} A\\ \hline G\end{array}\right),
$$
and that $[R^n,\cdots,R^n](GG^T)^{-1}=[R^n,\cdots,R^n]$.
By Theorem \ref{Thm3.2}, the dual code $C^\bot$ is as follows:
$$
C^\bot=[C_1^\bot,\cdots,C_m^\bot,R^n,\cdots,R^n]\cdot
\begin{pmatrix} D^{-1}\\ & (GG^T)^{-1}\end{pmatrix}
 \left(\begin{array}{c} A\\ \hline G\end{array}\right).
$$
Since $D^{-1}=\begin{pmatrix}u_1^{-1}\\&\ddots\\&&u_m^{-1}\end{pmatrix}$
and clearly $u_j^{-1}C_j^\bot=C_j^\bot$, for $j=1, \cdots , m$, we have
$$\begin{array}{rl}
\, & [C_1^\bot,\cdots,C_m^\bot,R^n,\cdots,R^n]\cdot
\begin{pmatrix} D^{-1}\\ & (GG^T)^{-1}\end{pmatrix}\\
= & \big[u_1^{-1}C_1^\bot,\,\cdots,u_m^{-1}C_m^\bot, R^n,\cdots,R^n] \\
= & [C_1^\bot,\cdots,C_m^\bot,R^n,\cdots,R^n],
\end{array}$$
so
\begin{eqnarray*}
C^\bot&=&[C_1^\bot,\cdots,C_m^\bot,R^n,\cdots,R^n]\cdot
 \left(\begin{array}{c} A\\ \hline G\end{array}\right)\\
 &=&[C_1^\bot,\cdots,C_m^\bot]A+{\rm M}_{n\times(l-m)}(R)G\\
 &\supseteq&[C_1^\bot,\cdots,C_m^\bot]A\supseteq[C_1,\cdots,C_m]A=C.
\end{eqnarray*}
The proof is now complete. \qed

\medskip
The following corollary follows immediately from Theorem \ref{Thm3.3}:

\begin{corollary}\label{Cor-p12}
Let $C_1,\cdots,C_m$ be self-dual linear codes over $R$ of length $n$
and let $A$ be a quasi-orthogonal $m\times m$ matrix over $R$.
Then $C=[C_1,\cdots,C_m]A$ is a self-dual code.  
\end{corollary}

\section{Strongly Full-Row-Rank Matrices}


Let $C$ be a non-zero code over $R$ of length $n$
and set $M=|C|$ to be the cardinality of $C$.
If $d_H(C)=1$, then $M\le |R|^{n-d_H(C)+1}$.
In particular, we have $M\le |R|^{n-d_H(C)+1}$ when $n=1$.
If $n>1$ and $d_H(C)>1$, by puncturing at the last coordinate,
we get an $(n-1,M, \ge d-1)$ code $C'$,
where $d=d_H(C)$, and by induction, we obtain that
$M\le|R|^{(n-1)-(d-1)+1}=|R|^{n-d+1}$.
By this well-known argument (e.g., see \cite{NS}), we have the following
{\em Singleton bound} for codes over the Frobenius ring $R$:
\begin{equation}\label{eq4.1}
 d_H(C)\le n-\log_{|R|}|C|+1.
\end{equation}
If a code $C$ over $R$ of length $n$ attains the Singleton bound,
i.e., the equality holds in (\ref{eq4.1}),
then we say that $C$ is a {\em maximum distance separable} code over $R$, or an
{\em MDS} code over $R$ for short. Note, in particular, that $C = R^n$ is an MDS code.
We also adopt the convention that the zero code
is an MDS code (this is consistent with the convention that $d_H(0) = n+1$).

Note that, if $C$ is a free code over $R$ of length $l$, then (\ref{eq4.1}) becomes
$$
 d_H(C)\le l-{\rm rank}(C)+1,
$$
and $C$ is MDS if and only if, for any non-zero codeword ${\bf c}\in C$,
we have $w_H({\bf c})>l-{\rm rank}(C)$.
Moreover, a free code of length $l$ and rank $m$, which we shall
also call an $[l,m]$ code (over $R$), has
FRR generator matrices of size $m\times l$.

The following result is well known for codes over finite fields (see, for example, 
\cite[Theorems 5.3.2 and 5.3.3]{R}).

\begin{lemma}\label{Lem4.1}
Let $A\in{\rm M}_{m\times l}(R)$ be FRR
and let $C=U_A(m)$ (i.e., $C$ is the free code over $R$ of length $l$ generated by the rows of $A$).
Then the following statements are equivalent:
\begin{itemize}
\item[(i)] $C$ is an $[l,m]$ MDS code.

\item[(ii)] Any $m\times m$ submatrix of $A$ is non-singular.

\item[(iii)] The dual code $C ^\bot$ of $C$ is an $[l,l-m]$ MDS code.
\end{itemize}
\end{lemma}

The proof of Lemma \ref{Lem4.1} is similar to that of \cite[Theorems 5.3.2 and 5.3.3]{R}. 
The analogous ingredients needed for our setting (over a finite commutative Frobenius ring) 
are found in Proposition \ref{Prop2.5} and Corollary \ref{Cor2.4}. 

\begin{remark}\label{Rem-p13}
{\rm
\begin{itemize}
\item[(i)] Note that there is another statement

$\bullet$~ {\it ``Any $(l-m)\times(l-m)$ submatrix of
a check matrix of $C$ is non-singular''}

\noindent which is equivalent to any of the three statements in Lemma \ref{Lem4.1},
but it is already indirectly covered by Lemma \ref{Lem4.1}.

\item[(ii)] If $C=R^l$, then $A$ is invertible and $C^\bot=0$. In this case, we adopt the convention
that the zero code is an MDS code with zero as a generator matrix.
Recall that we have also adopted the convention that $L_Q(l+1)=0$, for any $l\times l$ matrix $Q$.
\end{itemize} }
\end{remark}

In view of Lemma \ref{Lem4.1}, we introduce the following terminologies.

\begin{definition}\label{Def4.1}
Let $A$ be an FRR $m\times l$ matrix over $R$.
\begin{itemize}
\item[(i)] If $U_A(m)=L_A(1)$ is an $[l,m]$ MDS code,
then we say that $A$ is a {\em strongly full-row-rank (SFRR) matrix}.

\item[(ii)] For $t \ge 2$, if there is a sequence of indices
$0 = i_0 <i_1<\cdots<i_t=m$ such that
$U_A(i_h)$, for $h=0, 1,\cdots,t$, are MDS codes, then we say that
$A$ is an
{\em $(i_1,\cdots,i_{t-1})$-SFRR matrix}.
(When $t=1$, $A$ is just an SFRR matrix.)

\item[(iii)] For $t \ge 2$, if there is a sequence of
indices $1=i_0 < i_1 < \cdots<i_{t-1} < i_t = m+1$ such that
$L_A(i_h)$, for $h=0, 1,\cdots,t$, are MDS codes, then we say that
$A$ is a
{\em reversely $(i_1,\cdots,i_{t-1})$-SFRR matrix}. (When $t=1$, $A$ is just an
SFRR matrix.)
\end{itemize}
\end{definition}

\begin{proposition}\label{Prop4.2}
Let $A\in{\rm M}_{m\times l}(R)$ be FRR
and let $0= i_0 <i_1<\cdots<i_t=m$. Assume that
$\tilde A\in{\rm M}_{l\times l}(R)$ is an invertible matrix
with $A$ as the submatrix consisting of its first $m$ rows.
Then $A$ is an $(i_1,\cdots,i_{t-1})$-SFRR matrix if and only if
$(\tilde A^{-1})^T$ is a reversely $(i_1+1,\cdots,i_{t-1}+1, m+1)$-SFRR matrix
(or, if $m=l$, a reversely $(i_1+1,\cdots,i_{t-1}+1)$-SFRR matrix).
\end{proposition}

\noindent{\bf Proof.}~ Since $(\tilde A^{-1})^T$ is invertible,
$L_{(\tilde A^{-1})^T}(1) = R^l$.
Hence, $U_A(0)=0$, $L_{(\tilde A^{-1})^T}(1)$ and
$L_{(\tilde A^{-1})^T}(l+1)=0$ are all MDS codes.

Let $k=i_h$ with $1\le h\le t$.
It is enough to show that $U_A(k)=U_{\tilde A}(k)$ is an MDS code if and only if
$L_{(\tilde A^{-1})^T}(k+1)$ is an MDS code.

According to Proposition \ref{Prop2.6},
we write $\tilde A=\left(\begin{array}{c}A'\\ \hline A''\end{array}\right)$,
where $A'$ is the submatrix consisting of the first $k$ rows of $A$,
and write $\tilde A^{-1}=(B'\,|\,B'')$ correspondingly; then
$U_A(k)=U_{A'}(k)$ and $U_{A'}(k)^\bot=L_{B''^T}(1)=L_{(\tilde A^{-1})^T}(k+1)$.
Therefore, the proposition follows from Lemma \ref{Lem4.1} at once.
\qed

\medskip
Recall that a matrix $A=(a_{ij})_{m\times l}$ over $R$ is said to be
{\em non-singular by columns} if, for every $t$ with $1\le t\le m$, any $t\times t$
submatrix of the first $t$ rows of $A$ is non-singular.

From Lemma \ref{Lem4.1} and Proposition \ref{Prop4.2},
we have the following obvious consequence which is a generalization of
\cite[Proposition~7.2 and Theorem~6.6(i)]{Blackmore-Norton}.

\begin{corollary}\label{Cor4.3}
Let $A\in{\rm M}_{m\times l}(R)$ be FRR.
Assume that $\tilde A\in{\rm M}_{l\times l}(R)$ is an invertible matrix that
has $A$ as the submatrix of its first $m$ rows.
Then the following statements are equivalent:
\begin{itemize}
\item[(i)] $A$ is non-singular by columns.

\item[(ii)] $A$ is a $(1,2,\cdots,m-1)$-SFRR matrix.

\item[(iii)] $(\tilde A^{-1})^T$ is a reversely $(2,\cdots,m,m+1)$-SFRR matrix. (When
$m=l$, $(\tilde A^{-1})^T$ is a reversely $(2,\cdots,m)$-SFRR matrix.)
\end{itemize}
In particular, when $m=l$, the square matrix $A$ is non-singular by columns
if and only if $(A^{-1})^T$ is reversely non-singular by columns. 
\end{corollary}

\begin{example}\label{Ex4.1}
{\rm Let $T=\begin{pmatrix}1&0&1\\ 0&1&1\\ 1&1&1\end{pmatrix}$,
which is the matrix for the $(a+x|b+x|a+b+x)$-construction.
Then $T$ is a $(2)$-SFRR matrix, but $T$ is not non-singular by columns
because $U_T(1)$ is not MDS.
We note that $T$ is also a reversely $(3)$-SFRR matrix.

Observe also that $(T^{-1})^T=\begin{pmatrix}0&-1&1\\ -1&0&1\\ 1&1&-1\end{pmatrix}$
is a reversely $(3)$-SFRR matrix (cf. Proposition \ref{Prop4.2}).
}
\end{example}

The following lower bound is a generalization of
the main result of \cite{Blackmore-Norton}, and
the condition for the equality is a generalization of \cite[Theorem 1]{H-L-R}
to SFRR matrices over finite Frobenius rings.

\begin{theorem}\label{Thm4.5}
Let $A\in{\rm M}_{m\times l}(R)$ be an $(i_1,\cdots,i_{t-1})$-SFRR matrix,
where $0=i_0<i_1<\cdots<i_t=m$. Let $C_1,\cdots,C_m$ be codes over $R$ of length $n$
and let $C=[C_1,\cdots,C_m]A$. Then
\stepcounter{equation}\begin{equation}\tag{\theequation
U}\label{eq4.2U}
 d_H(C)\ge\min\big\{(l-i_h+1)d_H(C_{k_h}) \mid h=1,\cdots, t,~ i_{h-1}<k_h\le i_h \big\}.
\end{equation}
Furthermore, if the following three conditions are satisfied:
\begin{itemize}
\item[{\rm(E1)}] $C_1,\cdots,C_m$ are linear,

\item[{\rm(E2)}] $C_1=\cdots=C_{i_1}$, $C_{i_1+1}=\cdots=C_{i_2}$, $\cdots$,
$C_{i_{t-1}+1}=\cdots=C_{i_t}(=C_m)$,

\item[{\rm(E3)}] $C_{i_1}\supseteq C_{i_2}\supseteq \cdots \supseteq C_{i_t}$,
\end{itemize}
\noindent then equality holds in \eqref{eq4.2U}, i.e.,
\stepcounter{equation}\begin{equation}\tag{\theequation
U}\label{eq4.3U}
 d_H(C)=\min\big\{(l-i_h+1)d_H(C_{i_h}) \mid h=1,\cdots, t \big\}.
\end{equation}
\end{theorem}

\begin{remark}\label{Rem-p15}
{\rm There is a dual version of Theorem \ref{Thm4.5}, which we now state.
Let $A$ be a reversely $( i_1,\cdots,i_{t-1})$-SFRR $m\times l$ matrix over $R$,
where $1=i_0 < i_1<\cdots<i_{t-1}< i_t = m+1$.
Then the analogue of \eqref{eq4.2U} is:
\addtocounter{equation}{-1}
\begin{equation}\tag{\theequation L}\label{eq4.2L}
 d_H(C)\ge\min\big\{(l-m+i_h)d_H(C_{k_h}) \mid
  h=0, 1,\cdots, t-1,~ i_{h}\le k_h< i_{h+1} \big\}.
\end{equation}
With further conditions (E1$^*$)=(E1) and
\begin{itemize}
\item[(E2$^*$)] $(C_1=)C_{i_0}=\cdots=C_{i_1-1}$,
$C_{i_1}=\cdots=C_{i_2-1}$, $\cdots$, $C_{i_{t-1}}=\cdots=C_{m}$,
\item[(E3$^*$)] $C_{i_0}\subseteq C_{i_1}\subseteq \cdots \subseteq C_{i_{t-1}}$,
\end{itemize}
the analogous version of the equality \eqref{eq4.3U} is:
\stepcounter{equation}\begin{equation}\tag{\theequation L}\label{eq4.3L}
 d_H(C)=\min\big\{(l-m+i_h)d_H(C_{i_h}) \mid h=0, 1,\cdots, t-1 \big\}.
\end{equation}
The proof for the dual version is the same as that for Theorem \ref{Thm4.5}.
}
\end{remark}

\noindent{\bf Proof of Theorem \ref{Thm4.5}.}~
By Theorem \ref{Thm3.1} \eqref{eq3.3U}, we have that
$$
 d_H(C)\ge\min\big\{d_H(U_A(k))d_H(C_{k}) \mid 1\le k\le m \big\}.
$$
If $i_{h-1}<k\le i_h$, then $U_A(k)\subseteq U_A(i_h)$,
so $d_H(U_A(k))\ge d_H(U_A(i_h))=l-i_h+1$. Hence
$$ d_H(U_A(k))d_H(C_{k})\ge(l-i_h+1)d_H(C_{k}),\qquad i_{h-1}<k\le i_h.$$
The inequality  \eqref{eq4.2U} holds.

In order to prove \eqref{eq4.3U}, first we show that the following lemma holds.

\begin{lemma}\label{Lem4.6}
Let $A$ be as in Theorem \ref{Thm4.5} and
set $m_h=i_h-i_{h-1}$, for $h=1,\cdots,t$. Then
there is a block lower triangular matrix $Q$:
\begin{equation}\label{eq4.4}
Q=\begin{pmatrix}Q_1\\ *&Q_2\\\vdots&\ddots&\ddots\\ *&\cdots&*&Q_t\end{pmatrix},
\end{equation}
with $Q_h$ being an invertible $m_h\times m_h$ matrix for each $h=1,\cdots,t$,
such that $QA$ is a block upper triangular matrix
\begin{equation}\label{eq4.5}
QA=\begin{pmatrix} I_{m_1}& * &\cdots& * & \cdots & * \\
       & I_{m_2} & \cdots& *& \cdots & * \\
       & & \ddots &  \vdots & \vdots & \vdots \\
       &&& I_{m_t} & \cdots & * \end{pmatrix} ,
\end{equation}
where, for $h=1,\cdots,t$, the $i_h$th row of $QA$ takes the form
\begin{equation}\label{eq4.6}
\big(\underbrace{0,\;\cdots,\;0}_{i_h-1}\,,~
      1, u_{i_h,i_h+1},~ \cdots,~u_{i_h,l}\big)
\end{equation}
with $u_{i_h,j}$ being a unit of $R$ for every $j=i_h+1,\cdots,l$.
\end{lemma}

\noindent{\bf Proof.}~
Write $A=(a_{ij})_{m\times l}$, and consider the
top-left $i_1\times i_1$ submatrix $A_1=(a_{ij})_{i_1\times i_1}$.
By the assumption on $A$ and Lemma \ref{Lem4.1},
the submatrix $A_1$ is non-singular,
hence there is an $m_1\times m_1$ (recall that $m_1=i_1$)
invertible matrix $Q_1$ such that $Q_1A_1=I_{m_1}$. Setting
$$ Q'=\begin{pmatrix}Q_1\\ & I_{m_2} \\ &&\ddots\\ &&&I_{m_t}\end{pmatrix} , $$
it follows that
$$ Q'A=\begin{pmatrix} I_{m_1}&*&\cdots&*\\ *&*&\cdots&*
   \\ \vdots&\vdots&\vdots&\vdots\\ *&*&\cdots&*\end{pmatrix} . $$
By adding suitable multiples of the rows in the first row partition to the rows in the
other row partitions,
we obtain an invertible matrix
$$ Q''=\begin{pmatrix}Q_1\\ * & I_{m_2} \\ \vdots&&\ddots\\ *&&&I_{m_t}\end{pmatrix} $$
such that
$$ Q''A=\begin{pmatrix} I_{m_1}&*&\cdots&*\\ &*&\cdots&*
   \\ &\vdots&\vdots&\vdots\\ &*&\cdots&*\end{pmatrix}. $$
Note that, by the properties of determinants and Lemma \ref{Lem4.1},
$U_{Q''A}(i_h)$, for $h=1,\cdots,t$, are still MDS codes.
The top-left $i_2\times i_2$ submatrix of $Q''A$ looks like
$$
 \left(\begin{array}{c|c} I_{m_1} & * \\\hline  & A_2 \end{array}\right) ,
$$
which should be non-singular, hence $A_2$ is an invertible $m_2\times m_2$ matrix,
where $m_2=i_2-i_1$. Thus we can repeat the above process
until $Q$ satisfying conditions \eqref{eq4.4} and \eqref{eq4.5} is found.

Note that the $i_h$th row of $QA$ in \eqref{eq4.5} has the form of \eqref{eq4.6},
except that it remains to show that $u_{i_h,j}$, for all $j\ge i_{h}+1$, are units of $R$.
Consider the $i_h\times i_h$ submatrix of $QA$ formed by the
first $i_h$ rows and the $1$st, $2$nd, $\cdots$, $(i_h-1)$th and the $j$th columns:
$$
 \begin{pmatrix} 1&\cdots&*&*\\ &\ddots&\vdots&\vdots\\ &&1&*\\&&&u_{i_h,j} \end{pmatrix} .
$$
Since $U_{QA}(i_h)$ is still an MDS code, this submatrix is non-singular,
hence its determinant $u_{i_h,j}$ is a unit of $R$. \qed

\medskip With the notations in Lemma \ref{Lem4.6},
we return to the proof of Theorem \ref{Thm4.5}.
Note that $Q^{-1}=(r_{ij})_{m\times m}$ is also a block lower triangular matrix
$$Q^{-1}=(r_{ij})_{m\times m}=\begin{pmatrix}Q_1^{-1}\\ *&Q_2^{-1}\\
   \vdots&\ddots&\ddots\\ *&\cdots&*&Q_t^{-1}\end{pmatrix},$$
that is,
$$r_{ij}=0, \qquad i\le i_h < j,\quad h=1,\cdots,t-1. $$
Then
\begin{eqnarray*}
 C&=&[C_1,\cdots,C_m]A=[C_1,\cdots,C_m]Q^{-1}QA\\
  &=&[C_1,\cdots,C_m]Q^{-1} \cdot
   \begin{pmatrix} I_{m_1}& * &\cdots& * & \cdots & * \\
       & I_{m_2} & \cdots& *& \cdots & * \\
       & & \ddots &  \vdots & \vdots & \vdots \\
       &&& I_{m_t} & \cdots & * \end{pmatrix} .
\end{eqnarray*}
For any $({\bf c}_1,\cdots,{\bf c}_m)\in[C_1,\cdots,C_m]$,
write $({\bf c}_1,\cdots,{\bf c}_m)Q^{-1}=({\bf c}'_1,\cdots,{\bf c}'_m)$ with
$$
  {\bf c}'_k= r_{1k}{\bf c}_1+\cdots+r_{mk}{\bf c}_m .
$$
For $h=1, \cdots , t$ and  $i_{h-1}<k\le i_h$, since $r_{ik}=0$
for $i \le i_{h-1}$, we have
$${\bf c}'_k= r_{i_{h-1}+1,k}{\bf c}_{i_{h-1}+1}+r_{i_{h-1}+2,k}{\bf c}_{i_{h-1}+2}
+\cdots+r_{mk}{\bf c}_m;$$
so, by the conditions (E1),(E2) and (E3), we have that
${\bf c}'_k\in C_k$.
Hence, ${\bf c}'_k\in C_k$ for all $k=1,\cdots,m$, implying
$[C_1,\cdots,C_m]Q^{-1}\subseteq [C_1,\cdots,C_m]$.
Moreover, $Q^{-1}$ is an invertible matrix, so
$[C_1,\cdots,C_m]Q^{-1}=[C_1,\cdots,C_m]$.
Therefore,
$$
 C=[C_1,\cdots,C_m](QA)
   =[C_1,\cdots,C_m]\begin{pmatrix} I_{m_1}& * &\cdots& * & \cdots & * \\
       & I_{m_2} & \cdots& *& \cdots & * \\
       & & \ddots &  \vdots & \vdots & \vdots \\
       &&& I_{m_t} & \cdots & * \end{pmatrix}.
$$
By the inequality \eqref{eq4.2U} and the condition (E2),
\begin{equation}\label{eq4.7}
 d_H(C)\ge\min\big\{(l-i_h+1)d_H(C_{i_h}) \mid h=1,\cdots, t \big\}.
\end{equation}
To prove \eqref{eq4.3U}, it is enough to show that, for each $h$ with $1\le h\le t$,
there is some ${\bf c}\in C$ such that $w_H({\bf c})=(l-i_h+1)d_H(C_{i_h})$.
For this purpose, we take
${\bf c}_{i_h}=(c_1,\cdots,c_n)\in C_{i_h}$ such that $w_H({\bf c}_{i_h})=d_H(C_{i_h})$.
By \eqref{eq4.6}, we get a codeword ${\bf c}\in C$ as follows:
$$
 {\bf c}=({\bf 0},\cdots,{\bf 0},{\bf c}_{i_h},{\bf 0},\cdots,{\bf 0})(QA)
 =\big(\underbrace{{\bf 0},\;\cdots,\;{\bf 0}}_{i_h-1}\,,~
      {\bf c}_{i_h},\;u_{i_h,i_h+1}{\bf c}_{i_h},~
       \cdots,~u_{i_h,l}{\bf c}_{i_h}\big).
$$
Since $u_{i_h,j}$ are units for all $j\ge i_{h}+1$, it follows that
$w_H(u_{i_h,j}{\bf c}_{i_h})=w_H({\bf c}_{i_h})=d_H(C_{i_h})$. Hence, we obtain that
$$
 w_H({\bf c})=w_H({\bf c}_{i_h})+w_H(u_{i_h,i_h+1}{\bf c}_{i_h})+
  \cdots+w_H(u_{i_h,l}{\bf c}_{i_h})=(l-i_h+1)d_H(C_{i_h}),
$$
which completes the proof of Theorem \ref{Thm4.5}. \qed

\medskip
We next consider the analogue of Theorem \ref{Thm4.5} for the dual code.

Let $A\in{\rm M}_{m\times l}(R)$ be an $( i_1,\cdots,i_{t-1})$-SFRR matrix,
where $0=i_0<i_1<\cdots<i_t=m$.
Let $C_1,\cdots,C_m$ be codes over $R$ of length $n$ and let $C=[C_1,\cdots,C_m]A$.
From Theorem \ref{Thm3.2}, we recall that the dual code is
\begin{equation}\label{eq4.8}
 C^\bot=[C_1^\bot,\cdots,C_m^\bot,\,\underbrace{R^n,\cdots,R^n}_{l-m}\,]
 (\tilde A^{-1})^T,
\end{equation}
where $\tilde A\in{\rm M}_{l\times l}(R)$ is an invertible matrix
with $A$ as the submatrix consisting of its first $m$ rows
(see Remark \ref{Rem-p10}).
Now we estimate the minimum distance of $C^\bot$.
If $m<l$, we have $C_{m+1}^\bot=\cdots=C_l^\bot=R^n$ and
set $i_{t+1}=l$ for convenience.

\begin{theorem}\label{Thm4.7}
Let the notations be as above. Then
\begin{equation}\label{eq4.9}
 d_H(C^\bot)\ge\min\big\{
  (i_h+1)d_H(C_{k_h}^\bot) \mid h=0,1,\cdots, t,~ i_{h}<k_h\le i_{h+1} \big\}.
\end{equation}
Furthermore, if the following three conditions are satisfied:
\begin{itemize}
\item[{\rm(E1)}] $C_1,\cdots,C_m$ are linear,

\item[{\rm(E2)}] $C_1=\cdots=C_{i_1}$, $C_{i_1+1}=\cdots=C_{i_2}$, $\cdots$,
$C_{i_{t-1}+1}=\cdots=C_{i_t}$,

\item[{\rm(E3)}] $C_{i_1}\supseteq C_{i_2}\supseteq \cdots \supseteq C_{i_t}$,
\end{itemize}
then equality holds in \eqref{eq4.9}, i.e.,
\begin{equation}\label{eq4.10}
 d_H(C^\bot)=\min\big\{(i_h+1)d_H(C_{i_{h+1}}^\bot) \mid h=0,1,\cdots,t\big\}.
\end{equation}
\end{theorem}

\begin{remark}\label{Rem-p18}
{\rm If $m<l$ (i.e., $A$ is not square), then in the braces
of the right hand side of \eqref{eq4.9}, the terms for $h=t$ are:
$$(i_t+1)d_H(C_{k_t}^\bot)=m+1,\qquad i_t=m<k_t\le l=i_{t+1}. $$
Accordingly, in \eqref{eq4.10}, the term corresponding to $h=t$ is $m+1$.

On the other hand, when $m=l$, then in \eqref{eq4.9},
there is no term for $h=t$ since no $k$
satisfies  $l<k\le l$.
Accordingly, in \eqref{eq4.10}, there is no term $m+1$ for $h=t$. }
\end{remark}

\noindent{\bf Proof of Theorem \ref{Thm4.7}.}~ Let $\tilde B=\tilde A^{-1}$.
For $\tilde A$, we have that

$\bullet$~ $U_{\tilde A}(i_h)=U_{A}(i_h)$, for $h=0,1,\cdots,t$, are MDS codes.

\noindent By Proposition \ref{Prop4.2}, this is equivalent to

$\bullet$~ $L_{\tilde B^T}(i_h+1)$, for $h=0,1,\cdots,t$, are MDS codes.
(Note that $L_{\tilde B^T}(l+1)$ is trivially MDS.)

Since ${\rm rank}\big(L_{\tilde B^T}(i_h+1)\big)=l-i_h$,
we have that
$$d_H\big(L_{\tilde B^T}(i_h+1)\big)=l-(l-i_h)+1=i_h+1,  \qquad h=0,1,\cdots,t.$$
By the dual of Theorem \ref{Thm4.5} (see \eqref{eq4.2L}), we have that
$$
 d_H(C^\bot)\ge\min\big\{\,(i_h+1)d_H(C_{k_h}^\bot) \mid
   h=0,1,\cdots, t,~ i_{h}<k_h\le i_{h+1} \big\}.
$$
However, note that, if $i_t=m<l$, then, for any $k$ with $m<k\le l$,
we have that $C_k^\bot=R^n$,
hence $d_H(C_k^\bot)=1$; so the terms for $h=t$ in the braces are:
$$(i_t+1)d_H(C_{k_t}^\bot)=m+1,\qquad i_t=m<k_t\le l=i_{t+1}. $$
The inequality \eqref{eq4.9} is proved.

Further, assume that the conditions (E1), (E2) and (E3) hold.
Then, for the dual codes, the following conditions hold:
\begin{itemize}
\item[{\rm(E1$^*$)}] $C_1^\bot,\cdots,C_m^\bot$
are linear (note: $C_{m+1}^\bot, \cdots , C_l^\bot$ are trivially linear),

\item[{\rm(E2$^*$)}] $C_1^\bot=\cdots=C_{i_1}^\bot$,
$C_{i_1+1}^\bot=\cdots=C_{i_2}^\bot$, $\cdots$,
$C_{i_{t-1}+1}^\bot=\cdots=C_{i_t}^\bot$,
(note: $C_{m+1}^\bot = \cdots = C_l^\bot$ trivially),

\item[{\rm(E3$^*$)}] $C_{i_0+1}^\bot\subseteq C_{i_1+1}^\bot\subseteq \cdots
 \subseteq C_{i_{t-1}+1}^\bot\subseteq C_{i_t+1}^\bot =  R^n$.
\end{itemize}
By the dual of Theorem \ref{Thm4.5} (see \eqref{eq4.3L}),
we obtain the equality \eqref{eq4.10}.
(Note that, similar to the case of \eqref{eq4.9},
when $m < l$, the term corresponding to
$h=t$ is $m+1$, while, for the case $m=l$, there is no term $m+1$ for $h=t$.)
\qed

\medskip
As a special case, we have the following corollary
on non-singular by columns matrices over $R$,
which generalizes \cite[Theorems 3.7 and 6.6]{Blackmore-Norton} and
\cite[Propositions 2 and 4]{Van-Asch}.
However, in our case, for the bound on $d_H(C^\bot)$, we do
not require $A$ to be square.

\begin{corollary}\label{Cor4.8}
Let $A\in{\rm M}_{m\times l}(R)$ be non-singular by columns,
let $C_1,\cdots,C_m$ be codes over $R$ of length~$n$,
and let $C=[C_1,\cdots,C_m]A$. Then
$$d_H(C)\ge\min\big\{\,l\cdot d_H(C_1),\,(l-1)d_H(C_2),\,\cdots,\,
 (l-m+1)d_H(C_m)\,\big\}$$
and
$$d_H(C^\bot)\ge \left\{ \begin{array}{ll}
\min\big\{\, 1\cdot d_H(C_1^\bot),\,2\cdot d_H(C_2^\bot),\,
            \cdots,\,m\cdot d_H(C_m^\bot),\,m+1\,\big\}  & {\mbox{ if }} m < l , \\
\min\big\{\, 1\cdot d_H(C_1^\bot),\,2\cdot d_H(C_2^\bot),\,
            \cdots,\,m\cdot d_H(C_m^\bot)\,\big\}  & {\mbox{ if }} m = l . \end{array}
            \right.       $$

Further, if $C_1,\cdots,C_m$ are linear and $C_1\supseteq\cdots\supseteq C_m$,
then equalities are attained in all these inequalities. 
\end{corollary}

In the next section, we further discuss the properties of codes constructed 
with a special type of $(m')$-SFRR matrices, and provide two examples of codes constructed in 
this manner.

\section{Two-Way $(m')$-SFRR Matrices}


Recall that the well-known $(a+x|b+x|a+b+x)$-construction is associated with the
matrix $T=\begin{pmatrix}1&0&1\\ 0&1&1\\ 1&1&1\end{pmatrix}$
and the matrix product code $C=[C_1,C_1,C_2]T$.
We have seen in Example \ref{Ex4.1} that $T$ is a $(2)$-SFRR matrix
(but not a non-singular by columns matrix),
so \eqref{eq4.2U} of Theorem \ref{Thm4.5}
can be applied to show that the minimum distance
satisfies $d_H(C)\ge\min\{2d_H(C_1),d_H(C_2)\}$.
On the other hand, $T$ is also a reversely $(3)$-SFRR matrix, so
\eqref{eq4.2L} of Theorem \ref{Thm4.5}
is also applicable, yielding $d_H(C)\ge\min\{d_H(C_1),3d_H(C_2)\}$. Therefore,
$$d_H(C)\ge\max\big\{\min\{2d_H(C_1),d_H(C_2)\},\;
   \min\{d_H(C_1),3d_H(C_2)\}\big\}.$$
However, for this construction,
there is another well-known estimation (e.g., see \cite[Section V.B]{Forney}):
$$ \min\{d_H(C_1\cap C_2),\,2d_H(C_1),3d_H(C_2)\}\ge
d_H(C)\ge \min\{d_H(C_1\cap C_2),\,2d_H(C_1),3d_H(C_1+C_2)\}.
$$
Though the two lower bounds above cannot be directly compared in general,
in many cases the latter is better than the former.
Furthermore, we also note that $C$ is self-dual in many cases
though $T$ is not a quasi-orthogonal matrix.

Inspired by these observations, we introduce the following notion.

\begin{definition}\label{Def5.1}
Let $A\in{\rm M}_{m\times l}(R)$ be FRR.
If there is an index $m'$ with $1\le m'<m$ such that
$A$ is both an $(m')$-SFRR matrix and a reversely $(m'+1)$-SFRR matrix,
then we say that $A$ is a {\em two-way $(m')$-SFRR matrix}.
\end{definition}

\begin{remark}\label{Rem-p21}
{\rm For $m'+m''=m$,
any $m\times l$ matrix $A$ can be written as
$A=\left(\begin{array}{c}A'\\ \hline A''\end{array}\right)$,
where $A'$ is an $m'\times l$ matrix consisting of the first $m'$ rows of $A$
while $A''$ is an $m''\times l$ matrix consisting of the last $m''$ rows of $A$.
With this partitioned form, $A$ is a two-way $(m')$-SFRR matrix if and only if
$A'$, $A''$ and $A$ are all SFRR matrices.}
\end{remark}

The following property is a key point
for constructing self-orthogonal matrix product codes.

\begin{definition}\label{Def5.2}
Let an $m\times l$ matrix
$A=\left(\begin{array}{c}A'\\ \hline A''\end{array}\right)$
be partitioned into an $m'\times l$ matrix $A'$ and an $m''\times l$ matrix $A''$ as above.
If every row of $A'$ is orthogonal to every row of $A''$
with respect to the Euclidean inner product on $R^l$,
then we say that $A$ has a {\em partitioned orthogonal property},
or, more precisely, the {\em $m'$-partitioned orthogonal property}.
\end{definition}

\medskip
A quasi-orthogonal two-way $(m')$-SFRR matrix obviously has
the $m'$-partitioned orthogonal property.

\begin{example}\label{Ex5.1}
{\rm
\begin{itemize}
\item[(i)] As we have seen in Example \ref{Ex4.1}, for any Frobenius ring $R$,
$T=\begin{pmatrix}1&0&1\\ 0 & 1&1\\ 1&1&1\end{pmatrix}$ is a two-way $(2)$-SFRR matrix.
Furthermore, if $R$ has characteristic $2$,
then $T$ has the $2$-partitioned orthogonal property,
but $T$ is not quasi-orthogonal.
In fact, if $R$ is the binary field,
then $T$ is the unique two-way $(2)$-SFRR matrix of order $3$.

\item[(ii)] $A= \begin{pmatrix}1&1\\ 1&-1\end{pmatrix}$ is a two-way $(1)$-SFRR matrix provided
the characteristic of $R$ is different from~$2$. Moreover, $A$ is also a quasi-orthogonal matrix.

If $R$ is the binary field, then there are
no two-way $(1)$-SFRR matrices of order $2$ over $R$.
However, if $R$ is a field of characteristic $2$ but not the binary field,
taking any $1\ne\omega\in R$,
then $\begin{pmatrix}1&\omega\\ \omega&1\end{pmatrix}$ is a two-way $(1)$-SFRR matrix
which is also a quasi-orthogonal matrix.

\item[(iii)] $\begin{pmatrix}1&0&1&1\\ 0 & 1&1&-1\\ 1&1&1&0\\ 1&-1&0&1\end{pmatrix}$
is a two-way $(2)$-SFRR matrix if the characteristic ${\rm char}\,R\ne 2$.

However, if $3\nmid{\rm char}\,R$ and ${\rm char} \, R>2$, then
$\begin{pmatrix}1&0&1&1\\ 0 & 1&1&-1\\ 1&1&-1&0\\ 1&-1&0&-1\end{pmatrix}$
is a two-way $(2)$-SFRR matrix which is also a quasi-orthogonal matrix.

Note that, if $R$ is the binary field, there are no two-way
$(m')$-SFRR matrices over~$R$
of size $4\times 4$, for any $1 \le m' \le 3$.
\end{itemize} }
\end{example}

According to Remark \ref{Rem-p21},
we can partition a two-way $(m')$-SFRR matrix $A$ as
$A=\left(\begin{array}{c}A'\\ \hline A''\end{array}\right)$, where
$A'$ is an $m'\times l$ SFRR matrix and
$A''$ is an $m''\times l$ SFRR matrix.
For linear codes $C_1,\cdots,C_m$ over $R$ of length $n$, it is obvious that
the following two matrix product codes are equivalent to each other:
$$
\big[C_1,\cdots,C_{m'},C_{m'+1},\cdots,C_m\big]
 \left(\begin{array}{c}A'\\ \hline A''\end{array}\right)\cong
  \big[C_{m'+1},\cdots,C_m,C_{1},\cdots,C_{m'}\big]
   \left(\begin{array}{c}A''\\ \hline A'\end{array}\right).
$$
Without loss of generality, we can further assume that $m'\ge m''$.

\medskip
Let $A\in{\rm M}_{m\times l}(R)$, let $m'+m''=m$ with $m'\ge m''\ge 1$, and
let $C'$ and $C''$ be linear codes over $R$ of length $n$.
We consider the matrix product code
\begin{equation}\label{eq5.1}
C=
 [\,\underbrace{C',\cdots,C'}_{m'}\,,\,\underbrace{C'',\cdots,C''}_{m''}\,]A.
\end{equation}

If $A$ is a two-way $(m')$-SFRR matrix, then from \eqref{eq4.2U} and \eqref{eq4.2L}
of Theorem \ref{Thm4.5},
we have a lower bound for $d_H(C)$ as follows:
\begin{equation}\label{eq5.2}
 d_H(C)\ge\max\left\{\begin{array}{c}\min\{(l-m'+1)d_H(C'),\;(l-m+1)d_H(C'')\},\\[5pt]
    \min\{(l-m+1)d_H(C'),\;(l-m''+1)d_H(C'')\}\end{array}\right\}.
\end{equation}
Now we have some more bounds for $d_H(C)$ stated as follows.

\begin{theorem}\label{Thm5.1}
Let the notations be as in \eqref{eq5.1}.
If $A$ is a two-way $(m')$-SFRR matrix, then
\begin{equation}\label{eq5.3}
 d_H(C)\ge\min\big\{(l-m'+1)d_H(C'),\,(l-m''+1)d_H(C'+C''),\,(l-m+1)d_H(C'\cap C'')\big\}
\end{equation}
and
\begin{equation}\label{eq5.4}
 d_H(C)\le\min\big\{(l-m'+1)d_H(C'),\,(l-m''+1)d_H(C''),\,(l-m+1)d_H(C'\cap C'')\big\}.
\end{equation}
\end{theorem}

\noindent {\bf Proof.}~ Set $C_{\cap} = C'\cap C''$.
Since $[C',\cdots,C',\,C'',\cdots,C'']\supseteq[C',\cdots,C',\,C_\cap,\cdots,C_\cap]$, we have
$C=[C',\cdots,C',\,C'',\cdots,C'']A\supseteq[C',\cdots,C',\,C_\cap,\cdots,C_\cap]A$, so
$$d_H(C)\le d_H\big(
[\,\underbrace{C',\cdots,C'}_{m'}\,,\,\underbrace{C_{\cap},\cdots,C_{\cap}}_{m''}\,]A\big).
$$
Since $C'\supseteq C_\cap$, by \eqref{eq4.3U} of Theorem \ref{Thm4.5}, we have
\begin{equation}\label{eq5.5}
 d_H\big(
[\,\underbrace{C',\cdots,C'}_{m'}\,,\,\underbrace{C_{\cap},\cdots,C_{\cap}}_{m''}\,]A\big)
 =\min\big\{(l-m'+1)d_H(C'),\,(l-m+1)d_H(C_\cap)\big\},
\end{equation}
thus
\begin{equation}\label{eq5.6}
 d_H\big(C\big)\le\min\big\{(l-m'+1)d_H(C'),\,(l-m+1)d_H(C'\cap C'')\big\}.
\end{equation}
Applying \eqref{eq4.3L} to $C=[C', \cdots , C',C'', \cdots , C'']A
\supseteq[C_{\cap}, \cdots , C_{\cap},C'', \cdots , C'']A$ and observing that $C_\cap\subseteq C''$,
we obtain
\begin{equation}\label{eq5.7}
 d_H\big(C\big)\le\min\big\{(l-m''+1)d_H(C''),\,(l-m+1)d_H(C'\cap C'')\big\}.
\end{equation}
Combining \eqref{eq5.6} and \eqref{eq5.7},  the conclusion \eqref{eq5.4} follows.

Now we proceed to prove \eqref{eq5.3}. We partition $A$ as
$A=\left(\begin{array}{c}A'\\ \hline A''\end{array}\right)$,
where $A'$ is the $m'\times l$ matrix consisting of the first $m'$ rows of $A$
while $A''$ is the $m''\times l$ matrix consisting of the last $m''$ rows of $A$.
Assume that ${\bf c}'_1,\cdots,{\bf c}'_{m'}\in C'$,
${\bf c}''_1,\cdots,{\bf c}''_{m''}\in C''$ and
$\big({\bf c}'_1,\cdots,{\bf c}'_{m'},\,{\bf c}''_1,\cdots,{\bf c}''_{m''}\big)\ne{\bf 0}$.
We have a non-zero codeword of $C$ as follows:
$$
 {\bf c}=\big({\bf c}'_1,\cdots,{\bf c}'_{m'},\,{\bf c}''_1,\cdots,{\bf c}''_{m''}\big)A
 =\big({\bf c}'_1,\cdots,{\bf c}'_{m'}\big)A'+\big({\bf c}''_1,\cdots,{\bf c}''_{m''}\big)A''.
$$
We consider the $m''\times m''$ submatrices of $A''$: there are two cases. For any
matrix $M \in {\rm M}_{m\times l}(R)$ and $1 \le j_1 <
\cdots < j_s \le l$, let $M(j_1, \cdots , j_s)$ denote the $m \times s$ submatrix of $M$ consisting of the
$j_1$th, $\cdots$, $j_s$th columns of $M$.

\noindent {\bf Case 1:} There are $m''$ columns of $A$,
say the $j_1$th, $\cdots$, $j_{m''}$th columns, such that
$$
 \big({\bf c}'_1,\cdots,{\bf c}'_{m'}\big)A'(j_1,\cdots,j_{m''})
  +\big({\bf c}''_1,\cdots,{\bf c}''_{m''}\big)A''(j_1,\cdots,j_{m''})
 ={\bf 0}.
$$
Note that
$A''(j_1,\cdots,j_{m''})$ is an invertible $m''\times m''$ submatrix of $A''$
because $A''$ is an SFRR matrix. Then
$$
\big({\bf c}''_1,\cdots,{\bf c}''_{m''}\big)A''(j_1,\cdots,j_{m''})
=-\big({\bf c}'_1,\cdots,{\bf c}'_{m'}\big)A'(j_1,\cdots,j_{m''}),
$$
where the left hand side belongs to
$$
[\,\underbrace{C'',\cdots,C''}_{m''}\,]A''(j_1,\cdots,j_{m''})
  =[\,\underbrace{C'',\cdots,C''}_{m''}\,],
$$
and the right hand side belongs to
$$
[\,\underbrace{C',\cdots,C'}_{m'}\,]A'(j_1,\cdots,j_{m''})
  \subseteq[\,\underbrace{C',\cdots,C'}_{m''}\,].
$$
Thus
$$
\big({\bf c}''_1,\cdots,{\bf c}''_{m''}\big)A''(j_1,\cdots,j_{m''})
\in[\,\underbrace{C'',\cdots,C''}_{m''}\,]\bigcap[\,\underbrace{C',\cdots,C'}_{m''}\,]
=[\,\underbrace{C_{\cap},\cdots,C_{\cap}}_{m''}\,].
$$
It then follows that
$$
\big({\bf c}''_1,\cdots,{\bf c}''_{m''}\big)
\in [\,\underbrace{C_{\cap},\cdots,C_{\cap}}_{m''}\,]A''(j_1,\cdots,j_{m''})^{-1}
=[\,\underbrace{C_{\cap},\cdots,C_{\cap}}_{m''}\,].
$$
Hence,
$$
 {\bf c}=\big({\bf c}'_1,\cdots,{\bf c}'_{m'},\,{\bf c}''_1,\cdots,{\bf c}''_{m''}\big)A
 \in[\,\underbrace{C',\cdots,C'}_{m'}\,,\,\underbrace{C_{\cap},\cdots,C_{\cap}}_{m''}\,]A,
$$
and by \eqref{eq5.5}, we get
\begin{equation}\label{eq5.8}
 w_H({\bf c})\ge\min\big\{(l-m'+1)d_H(C'),\;(l-m+1)d_H(C' \cap C'')\big\} .
\end{equation}

\noindent {\bf Case 2:} There are at most $m''-1$ columns of $A$,
say the first $s$ columns, where $s \le m''-1$, such that
$$
 \big({\bf c}'_1,\cdots,{\bf c}'_{m'}\big)A'(1,\cdots,s)
  +\big({\bf c}''_1,\cdots,{\bf c}''_{m''}\big)A''(1,\cdots,s)={\bf 0}.
$$
By the construction \eqref{eq5.1} of $C$, ${\bf c}$ is an $n\times l$ matrix.
The above assumption means that 
any one of the last $l-m''+1$ columns of ${\bf c}$ is a non-zero vector of $R^n$.
By the construction \eqref{eq5.1} of $C$, any column of ${\bf c}$ is a vector of $C'+C''$, so
\begin{equation}\label{eq5.9}
 w_H({\bf c})\ge (l-m''+1)d_H(C'+C'').
\end{equation}

Summarizing the discussions for the two cases,
we see that, for any non-zero codeword ${\bf c}$ of $C$, one of \eqref{eq5.8}
and \eqref{eq5.9} holds, so we obtain
$$
 d_H(C)\ge\min\big\{(l-m'+1)d_H(C'),\,(l-m+1)d_H(C' \cap C''),\,(l-m''+1)d_H(C'+C'')\big\},
$$
which is just the required inequality \eqref{eq5.3}.
\qed

\begin{remark}\label{Rem-p24}
{\rm
\begin{itemize}
\item[(i)] For the proof of \eqref{eq5.3},
if we start with considering the $m'\times m'$ submatrices of~$A'$,
then we can obtain in a similar way that
\begin{equation}\label{eq5.10}
 d_H(C)\ge\min\big\{(l-m''+1)d_H(C''),\,(l-m+1)d_H(C'\cap C''),\,(l-m'+1)d_H(C'+C'')\big\}.
\end{equation}
Observing that $l-m'+1\le l-m''+1$ since we have assumed that $m'\ge m''$,
and that $d_H(C'+C'')\le d_H(C')$, we have that
$$
(l-m'+1)d_H(C'+C'')\le\min\{(l-m''+1)d_H(C'+C''),\,(l-m'+1)d_H(C')\},
$$
so
$$\begin{array}{c}
\min\big\{(l-m''+1)d_H(C''),\,(l-m+1)d_H(C'\cap C''),\,(l-m'+1)d_H(C'+C'')\big\}\\
\le\min\big\{(l-m'+1)d_H(C'),\,(l-m+1)d_H(C'\cap C''),\,(l-m''+1)d_H(C'+C'')\big\}.
\end{array}$$
In other words, under the assumption that $m'\ge m''$,
the bound \eqref{eq5.10} is not better than that of \eqref{eq5.3}.

\item[(ii)] However, the bounds in \eqref{eq5.2} and \eqref{eq5.3}
cannot be compared in general, because
$d_H(C'')$ in \eqref{eq5.2} and $(l-m''+1)d_H(C'+C'')$ in \eqref{eq5.3}
are not comparable in general.
Thus, we can take the larger of \eqref{eq5.2} and \eqref{eq5.3}
as a better lower bound for $d_H(C)$.
\end{itemize}
}
\end{remark}

\begin{theorem}\label{Thm5.2}
Let the notations be as in \eqref{eq5.1}.
Further assume that the matrix $A$ has the $m'$-partitioned orthogonal property.
If both $C'$ and $C''$ are self-orthogonal, then $C$ is self-orthogonal too.
In particular, $C$ is self-dual provided both $C'$ and $C''$ are self-dual
and $A$ is invertible.
\end{theorem}

\noindent{\bf Proof.}~
Write $A=\left(\begin{array}{c}A'\\ \hline A''\end{array}\right)$
with $A'$ being the $m'\times l$ matrix consisting of the first $m'$ rows of $A$
and $A''$ the $m''\times l$ matrix consisting of the last $m''$ rows of $A$.
By the product of partitioned matrices,
$$
 AA^T=\left(\begin{array}{c}A'\\ \hline A''\end{array}\right)
  \big(\,A'^T\mid A''^T\,\big)=
  \begin{pmatrix} A'A'^T & A'A''^T\\ A''A'^T&A''A''^T\end{pmatrix}.
$$
By the $m'$-partitioned orthogonal property, we have that $A'A''^T=0$ and $A''A'^T=0$, so
$$
 AA^T=\begin{pmatrix} A'A'^T \\ &A''A''^T\end{pmatrix}.
$$
Then, for any two codewords
$${\bf c}=\big({\bf c}'_1,\cdots,{\bf c}'_{m'},\,{\bf c}''_1,\cdots,{\bf c}''_{m''}\big)A\in C,
\quad
{\bf d}=\big({\bf d}'_1,\cdots,{\bf d}'_{m'},\,{\bf d}''_1,\cdots,{\bf d}''_{m''}\big)A\in C,
$$
with ${\bf c}'_1,\cdots,{\bf c}'_{m'},\,{\bf d}'_1,\cdots,{\bf d}'_{m'}\in C'$
and ${\bf c}''_1,\cdots,{\bf c}''_{m''},\,{\bf d}''_1,\cdots,{\bf d}''_{m''}\in C''$,
by \eqref{eq3.1}, we have
\begin{eqnarray*}
&&\langle{\bf c},{\bf d}\rangle={\rm tr}\big({\bf c}{\bf d}^T\big)
={\rm tr}\left(\big({\bf c}'_1,\cdots,{\bf c}'_{m'},\,{\bf c}''_1,\cdots,{\bf c}''_{m''}\big)A
 A^T\big({\bf d}'_1,\cdots,{\bf d}'_{m'},\,{\bf d}''_1,\cdots,{\bf d}''_{m''}\big)^T\right)\\
&=&{\rm tr}\left(\big(({\bf c}'_1,\cdots,{\bf c}'_{m'})(A'A'^T),\,
  ({\bf c}''_1,\cdots,{\bf c}''_{m''})(A''A''^T)\big)\cdot
  \big({\bf d}'_1,\cdots,{\bf d}'_{m'},\,{\bf d}''_1,\cdots,{\bf d}''_{m''}\big)^T\right).
\end{eqnarray*}
However, $ [C',\cdots,C'](A'A'^T)\subseteq [C',\cdots,C']$, so
$$({\bf c}'_1,\cdots,{\bf c}'_{m'})(A'A'^T)=(\bar{\bf c}'_1,\cdots,\bar{\bf c}'_{m'}),\qquad
   {\rm with}~~ \bar{\bf c}'_1,\cdots,\bar{\bf c}'_{m'}\in C'.
$$
Similarly,
$$({\bf c}''_1,\cdots,{\bf c}''_{m''})(A''A''^T)=(\bar{\bf c}''_1,\cdots,\bar{\bf c}''_{m''}),\qquad
   {\rm with}~~ \bar{\bf c}''_1,\cdots,\bar{\bf c}''_{m''}\in C''.
$$
Since both $C'$ and $C''$ are self-orthogonal,
\begin{eqnarray*}
\langle{\bf c},{\bf d}\rangle&=&{\rm tr}\left(
 \big(\bar{\bf c}'_1,\cdots,\bar{\bf c}'_{m'},\,\bar{\bf c}''_1,\cdots,\bar{\bf c}''_{m''}\big)
 \cdot\big({\bf d}'_1,\cdots,{\bf d}'_{m'},\,{\bf d}''_1,\cdots,{\bf d}''_{m''}\big)^T\right)\\
 &=&{\rm tr}\big(\bar{\bf c}'_1{\bf d}_1'^T\big)+\cdots+
       {\rm tr}\big(\bar{\bf c}'_{m'}{\bf d}_{m'}'^T\big)+
       {\rm tr}\big(\bar{\bf c}''_1{\bf d}_1''^T\big)+\cdots+
       {\rm tr}\big(\bar{\bf c}''_{m''}{\bf d}_{m''}''^T\big)\\
 &=&\langle\bar{\bf c}'_1,{\bf d}_1'\rangle+\cdots+\langle\bar{\bf c}'_{m'},{\bf d}_{m'}'\rangle+
  \langle\bar{\bf c}''_1,{\bf d}_1''\rangle+\cdots+\langle\bar{\bf c}''_{m''},{\bf d}_{m''}''\rangle\\
 &=& 0.
\end{eqnarray*}
Therefore, $C$ is self-orthogonal.

Assume that both $C'$ and $C''$ are self-dual.
Since $|C'||C'^\bot|=|R|^n$ and $|C''||C''^\bot|=|R|^n$, it follows that
$$|C'|=|C'^\bot|=|R|^{n/2}=|C''|=|C''^\bot|.$$
When $A$ is invertible, we have $m =l$,  so
$$
 |C|=|C'|^{m'}|C''|^{m''}=|R|^{(m'+m'')n/2}=|R|^{mn/2} = |R|^{ln/2}.
$$
Furthermore, from $|C||C^\bot|=|R|^{ln}$,
we have $|C^\bot|=|R|^{ln/2}$.
Since $C\subseteq C^\bot$, it
follows that $C=C^\bot$. \qed

\begin{example}\label{Ex5.2}
{\rm Take $R$ to be the binary field and
$T=\begin{pmatrix}1&0&1\\ 0&1&1\\ 1&1&1\end{pmatrix}$. Then $T$ is a
two-way $(2)$-SFRR matrix
that has the $2$-partitioned orthogonal property.
Recall that the matrix product code construction $C=[C',C',C'']T$ in \eqref{eq5.1} is just the well-known
$(a+x|b+x|a+b+x)$ construction. It is also a quasi-cyclic code
of co-index $3$ (see \cite[Theorem 6.7]{LS-I}).
The bounds in \eqref{eq5.3} and \eqref{eq5.4} of Theorem \ref{Thm5.1} give
the following well-known estimation on the minimum distance of $C$
(cf. \cite[Section V.B]{Forney}):
$$ \min\{d_H(C'\cap C''),\,2d_H(C'),3d_H(C'')\}\ge
d_H(C)\ge \min\{d_H(C'\cap C''),\,2d_H(C'),3d_H(C'+C'')\}.
$$
Another lower bound is given by \eqref{eq5.2}:
$$d_H(C)\ge\max\big\{\min\{2d_H(C'),d_H(C'')\},\;\min\{d_H(C'),3d_H(C'')\}\big\}.$$
It was noted in Remark \ref{Rem-p24} that these two lower bounds cannot be
compared directly in general. We now consider a few explicit examples.
First, we set

$$\begin{array}{c|c|c|c} \mbox{Code} & \mbox{parameters}
  & \mbox{generator matrix} & \mbox{duality}\\[1mm]\hline
C_1 & [4,1,4] & (1,1,1,1) &\mbox{self-orthogonal} \\[1mm]\hline
C_2 & [4,2,2] & \begin{pmatrix} 1 & 0 & 1 & 0\\ 0 & 1 & 1 & 1\end{pmatrix}
  & \mbox{not self-orthogonal}\\ \hline
C_3 & [4,2,2] & \begin{pmatrix} 1 & 0 & 1 & 0\\ 0 & 1 & 0 & 1\end{pmatrix}
  & \mbox{Type I self-dual} \\ \hline
C_3' & [4,2,2] & \begin{pmatrix} 1 & 1 & 0 & 0\\ 0 & 0 & 1 & 1\end{pmatrix}
  & \mbox{Type I self-dual} \\ \hline
\end{array}$$

\begin{itemize}
\item[(i)] Take $C=[C_2,C_2,C_1]T$, with $C_1, C_2$ as above.
Since $C_2\cap C_1=0$ and $C_2+C_1$
is a $[4,3,1]$ linear code with generator matrix
$\begin{pmatrix} 1 & 0 & 1 & 0\\ 0 & 1 & 1 & 1\\ 0 & 0 & 1 & 0\end{pmatrix}$,
the bound \eqref{eq5.2} shows that $d_H(C)\ge 4$, while the bound \eqref{eq5.3} gives $d_H(C)\ge 3$.
Therefore, in this case, the bound \eqref{eq5.2} is better than the bound \eqref{eq5.3}.
On the other hand, \eqref{eq5.4} shows that $d_H(C)\le 4$, hence, $d_H(C)=4$.
Thus $C$ is a $[12,5,4]$ binary linear code.
It can be verified directly that $C$ is not self-orthogonal. In fact, the following two codewords
are not orthogonal to each other:
\[ \left( \begin{array}{ccc} 0 & 0 & 1 \\ 0 & 1 & 1 \\ 0 & 1 & 1 \\ 0 & 1 & 1 \end{array}
\right) T = \left( \begin{array}{ccc} 1 & 1 & 1 \\ 1 & 0 & 0 \\ 1 & 0 & 0 \\ 1 & 0 & 0 \end{array}
 \right) \quad {\mbox{ and }} \quad
\left( \begin{array}{ccc} 1 & 0 & 1 \\ 0 & 0 & 1 \\ 1 & 0 & 1 \\ 0 & 0 & 1 \end{array} \right) T =
\left( \begin{array}{ccc} 0 & 1 & 0 \\ 1 & 1 & 1 \\ 0 & 1 & 0 \\ 1 & 1 & 1 \end{array} \right) .
\]

\item[(ii)] Take $C=[C_3,C_3,C_3']T$ with $C_3, C_3'$ as above.
Since $C_3\cap C_3'=C_1$ and $C_3+C_3'$
is a $[4,3,2]$ linear code with generator matrix
$\begin{pmatrix} 1 & 0 & 1 & 0\\ 0 & 1 & 0 & 1\\ 0 & 0 & 1 & 1\end{pmatrix}$.
The bound \eqref{eq5.2} shows that $d_H(C)\ge 2$,
while the bound \eqref{eq5.3} gives  $d_H(C)\ge 4$.
Hence, in this case, the bound \eqref{eq5.2} is weaker than the bound \eqref{eq5.3}.
From \eqref{eq5.4}, we obtain $d_H(C)\le 4$, thus $d_H(C)=4$.
Further, both $C'$ and $C''$ are self-dual in this case.
Therefore, by Theorem \ref{Thm5.2}, $C$ is a self-dual $[12,6,4]$ binary linear code.
However, $C$ is not of Type II: this follows from \cite[Proposition 7.1]{LS-I} with the
fact that $C_3'$ is not of Type II, but it can also be seen directly that the
codeword
\[ \left( \begin{array}{ccc} 1 & 0 & 1 \\ 0 & 1 & 1 \\ 1 & 0 & 0 \\ 0 & 1 & 0 \end{array}
\right) T = \left( \begin{array}{ccc} 0 & 1 & 0 \\ 1 & 0 & 0 \\ 1 & 0 & 1 \\
0 & 1 & 1 \end{array} \right)
\]
does not have Hamming weight divisible by 4.

\item[(iii)] Take $C=[C_3,C_3,C_1]T$ with $C_1, C_3$ as above.
Since $C_3\supseteq C_1$,
by \eqref{eq4.3U} we have that $d_H(C)=\min\{2d_H(C_3),\,d_H(C_1)\}=4$.
Hence, $C$ is a $[12,5,4]$ binary linear code.
Since $C_3$ is self-dual and $C_1$ is self-orthogonal,
$C$ is also self-orthogonal.

\medskip We summarize the above examples in the following:
$$\begin{array}{c|c|c|c} \mbox{Code $C$} & \mbox{parameters}
   & \mbox{duality} & \mbox{argument for $d_H(C)$}\\[1mm]\hline
[C_2,C_2,C_1]T & [12,5,4] & \mbox{not self-orthogonal}
  & \mbox{by \eqref{eq5.2}} \\[1mm]\hline
[C_3,C_3,C_3']T & [12,6,4] & \mbox{Type I self-dual}
  & \mbox{by \eqref{eq5.3}}\\ \hline
[C_3,C_3,C_1]T & [12,5,4] & \mbox{self-orthogonal}
  & \mbox{by \eqref{eq4.3U}} \\ \hline
\end{array}$$
\end{itemize}
}
\end{example}

\begin{example}\label{Ex5.3}
{\rm  Take $R$ to be the binary field. Take
$A=\begin{pmatrix}1&1\\ &1&1\\ &&1&1\\&&&1&1\\1&1&1&1&1\end{pmatrix}$,
which is a two-way $(4)$-SFRR matrix
that has the $4$-partitioned orthogonal property.
In fact, $A$ is the matrix for constructing quasi-cyclic codes of co-index 5,
see \cite[Theorem 6.14]{LS-I}.
Similar to the construction of the $[24,12,8]$-Golay code from
$[8,4,4]$-extended Hamming codes by the $(a+x|b+x|a+b+x)$-construction,
we construct $C=[C',C',C',C',C'']A$, where
$C'$ and $C''$ are $[8,4,4]$ extended Hamming codes with
generator matrices $G'$ and $G''$, respectively, as follows:
$$
 G'=\begin{pmatrix}1&1&0&1&&&&1\\&1&1&0&1&&&1\\&&1&1&0&1&&1\\&&&1&1&0&1&1\end{pmatrix},\qquad
 G''=\begin{pmatrix}1&0&1&1&&&&1\\&1&0&1&1&&&1\\&&1&0&1&1&&1\\&&&1&0&1&1&1\end{pmatrix}.
$$
It is known that both $C'$ and $C''$ are of Type II.
Since $C'\cap C''$ is an $[8,1,8]$ code and $C'+C''$ is an $[8,7,2]$ code,
by \eqref{eq5.3} and \eqref{eq5.4} we have that
$$
  8=\min\{2\cdot 4,\;5\cdot 2,\;1\cdot 8\}\le d_H(C)
   \le\min\{2\cdot 4,\;5\cdot 4,\;1\cdot 8\}=8,
$$
that is, $d_H(C)=8$. By Theorem \ref{Thm5.2}, $C$ is self-dual.
Furthermore, since $C''$ is of Type II, so is $C$ (see \cite[Proposition 7.3]{LS-I}).
We conclude that $C$ is a $[40,20,8]$ Type II binary code.
}
\end{example}

\section*{Acknowledgements}
This work was done while the first and third authors
were visiting the Division of Mathematical Sciences, School of Physical and Mathematical Sciences,
Nanyang Technological University, Singapore, in Autumn 2011.
They are grateful for the hospitality and support.
They also thank NSFC for the support through Grants No.~10871079 and No.~11171370.
The work of S.~Ling was partially supported
by Singapore MOE-AcRF Tier 2 Research
Grant T208B2204.

It is also the authors' pleasure to thank the anonymous referees for their 
helpful comments.


\begin{thebibliography}{99}

\smallskip

\bibitem{Blackmore-Norton} T. Blackmore and G. H. Norton,
{\em Matrix-product codes over $\mathbb{F}_q$},
Appl. Algebra Engrg. Comm. Comput., {\bf 12} (2001), 477--500.

\bibitem{Forney}G. D. Forney, {\em Coset codes II: binary lattices},
IEEE Trans. Inform. Theory, {\bf 34} (1988), 1152--1187.

\bibitem{Hammons} A. R. Hammons, P.V. Kumar, A. R. Calderbank,
 N. J. A. Sloane, and P. Sol\'e,
{\em The ${\bf Z}_4$-linearity of Kerdock, Preparata,
 Goethals, and related codes},
IEEE Trans. Inform. Theory, {\bf 40} (1994), 301--319.

\bibitem{H-H-R}F. Hernando, H. H{\o}holdt and D. Ruano, {\em 
List decoding of matrix-product codes from nested codes: an application 
to quasi-cyclic codes}, 2012, http://arxiv.org/pdf/1201.6397.pdf.

\bibitem{H-L-R}F. Hernando, K. Lally and D. Ruano,
{\em Construction and decoding of matrix-product codes from nested codes},
Appl. Algebra Engrg. Comm. Comput., {\bf 20} (2009), 497--507.

\bibitem{H-R-10} F. Hernando and D. Ruano,  {\em New linear codes from
matrix-product codes with polynomial units}, Adv. Math. Commun.,
{\bf 4} (2010), 363--367.

\bibitem{H-R-11} F. Hernando and D. Ruano,
{\em Decoding of matrix-product codes}, 2011,
http://arxiv.org/pdf/1107.1529.pdf.

\bibitem{LS-I} S. Ling and P. Sol\'e,
{\em On the algebraic structure of quasi-cyclic codes I: finite fields},
IEEE Trans. Inform. Theory, {\bf 47} (2001), 2751--2760.

\bibitem{LS-II} S. Ling and P. Sol\'e,
{\em On the algebraic structure of quasi-cyclic codes II: chain rings},
Des., Codes and Crypto., {\bf 30} (2003), 113--130.

\bibitem{M-M} E. Mart\'inez-Moro, {\em A generalization of Niederreiter-Xing's
propagation rule and its commutativity with duality}. IEEE Trans.
Inform. Theory, {\bf 50} (2004), 701--702.

\bibitem{Niederreiter-Xing}  H. Niederreiter and C. P. Xing,
{\em A propagation rule for linear codes},
Appl. Algebra Engrg. Comm. Comput., {\bf 10} (2000), 425--432.

\bibitem{NS} G. H. Norton and A. S\u{a}l\u{a}gean, {\em
On the Hamming distance of linear codes over a finite chain ring},
IEEE Trans. Inform. Theory, {\bf 46} (2000), 1060--1067.

\bibitem{Ould-Mamoun} M. B. O. Medeni and E. M. Souidi,
{\em Construction and bound on the performance of matrix-product codes},
Appl. Math. Sci. (Ruse), {\bf 5} (2011), 929--934.

\bibitem{Ozbudak-Stichtenoth}  F. \"{O}zbudak and H. Stichtenoth,
{\em Note on Niederreiter-Xing's propagation rule for linear codes},
Appl. Algebra Engrg. Comm. Comput., {\bf 13} (2002), 53--56.

\bibitem{R} S. Roman, {\em Coding and Information Theory}, Graduate 
Texts in Mathematics {\bf 134}, Springer-Verlag, 
New York, 1992. 

\bibitem{Van-Asch}  B. van Asch,
{\em Matrix-product codes over finite chain rings},
Appl. Algebra Engrg. Comm. Comput., {\bf 19} (2008), 39--49.

\bibitem{W-99} J. Wood,
{\em Duality for modules over finite rings and applications to coding theory},
Amer. J. Math., {\bf 121} (1999), 555--575.

\bibitem{W-08} J. Wood,
{\em Code equivalence characterizes finite Frobenius rings},
Proc. Amer. Math. Soc., {\bf 136} (2008), 699--706.

\end {thebibliography}
\end{document}